\documentclass[fleqn,usenatbib]{mnras}

\usepackage{newtxtext,newtxmath}
\usepackage[T1]{fontenc}

\usepackage{graphicx}
\usepackage{amsmath}
\usepackage{xspace}
\usepackage{bm}
\usepackage{ae,aecompl,times}
\usepackage{xpatch}
\usepackage{upgreek}
\usepackage{orcidlink}
\usepackage{hyperref}

\makeatletter
\xpatchcmd\NAT@citex
{%
	\@citea\NAT@hyper@{%
		\NAT@nmfmt{\NAT@nm}%
		\hyper@natlinkbreak{\NAT@aysep\NAT@spacechar}{\@citeb\@extra@b@citeb}%
		\NAT@date
	}%
}
{%
	\@citea
	\NAT@nmfmt{\NAT@nm}%
	\NAT@aysep\NAT@spacechar
	\NAT@hyper@{\NAT@date}%
}
{}{}
\xpatchcmd\NAT@citex
{%
	\@citea\NAT@hyper@{%
		\NAT@nmfmt{\NAT@nm}%
		\hyper@natlinkbreak{\NAT@spacechar\NAT@@open\if*#1*\else#1\NAT@spacechar\fi}%
		{\@citeb\@extra@b@citeb}%
		\NAT@date
	}%
}
{
	\@citea
	\NAT@nmfmt{\NAT@nm}%
	\NAT@spacechar\NAT@@open\if*#1*\else#1\NAT@spacechar\fi
	\NAT@hyper@{\NAT@date}%
}
{}{}
\makeatother
\newcommand{\Arepo}{\textsc{Arepo}\xspace}
\newcommand{\ArepoDG}{\textsc{Arepo-dg}\xspace}
\newcommand{\Tenet}{\textsc{Tenet}\xspace}

\newcommand{\ArepoCT}{\Arepo-CT\xspace}
\newcommand{\ArepoPowell}{\Arepo-Powell\xspace}
\newcommand{\ArepoDedner}{\Arepo-Dedner\xspace}

\newcommand{\DGPowell}{DG-Powell\xspace}
\newcommand{\DGDedner}{DG-Dedner\xspace}

\newcommand{\Ramses}{\textsc{Ramses}\xspace}
\newcommand{\RamsesCT}{\Ramses-CT\xspace}

\newcommand{\Athena}{\textsc{Athena}\xspace}
\newcommand{\AthenaPP}{\textsc{Athena++}\xspace}
\newcommand{\AthenaCT}{\Athena-CT\xspace}
\newcommand{\AthenaCTLR}{\Athena-CT-LR\xspace}
\newcommand{\AthenaCTPPM}{\Athena-CT-PPM\xspace}

\renewcommand{\vec}[1]{ {\bmath #1} } 

\title[Comparing driven subsonic MHD turbulence]{A comparison of numerical schemes for driven subsonic MHD turbulence}

\author[R. Pakmor et al.]{R\"udiger Pakmor$^{1}$, Thomas Guillet$^{2}$, Rebekka Bieri$^{3}$, Christoph Pfrommer$^{4}$, Volker Springel$^{1}$, and \newauthor%
Romain Teyssier$^{5}$\vspace*{0.1cm}\\%
$^{1}$Max Planck Institut f\"ur Astrophysik, Karl-Schwarzschildstr.~1, D-85748 Garching, Germany\\%
$^{2}$University of Exeter, Department of Physics and Astronomy, Exeter EX4 4QL, UK\\%
$^{3}$Institut f\"ur Astrophysik, Universit\"at Z\"urich, Winterthurerstrasse 190, 8057 Z\"urich, Switzerland\\%
$^{4}$Leibniz-Institut f{\"u}r Astrophysik Potsdam (AIP), An der Sternwarte 16, D-14482
Potsdam, Germany\\%
$^{5}$Department of Astrophysical Sciences, Princeton University, 4 Ivy Lane, Princeton NJ 08544, USA
}

\date{Accepted XXX. Received YYY; in original form ZZZ}

\pubyear{2026}

\begin{document}
\label{firstpage}
\pagerange{\pageref{firstpage}--\pageref{lastpage}}
\maketitle

\begin{abstract}
Turbulence is ubiquitous in astrophysical systems, and since most cosmic gas is ionised, it supports magnetic fields. In turbulent environments, these fields are rapidly amplified through a small-scale dynamo. Multi-scale astrophysical simulations, however, rarely resolve this process adequately. Limited spatial dynamic range makes small-scale amplification sensitive to the numerical choices made in the hydrodynamics and magnetic field solvers. Here, we investigate idealised periodic boxes of driven subsonic turbulence with a weak seed magnetic field. These simulations with purely numerical dissipation provide a simple environment in which an efficient small-scale dynamo is expected. We aim to systematically compare the three widely used magnetohydrodynamics (MHD) codes \textsc{Arepo}, \textsc{Athena}, and \textsc{Ramses} across different divergence-control schemes: constrained transport, Powell cleaning, and Dedner cleaning. To minimise comparison bias, we adopt identical turbulent driving and analysis pipelines across all runs. At sufficient resolution, every code and scheme we test exhibits dynamo-like exponential amplification of the seed field until saturation. The structural properties of the magnetic field in both the kinematic and saturated regimes are consistent across schemes (with the exception of \textsc{Arepo}'s constrained transport) and agree with theoretical expectations. Residual differences, particularly in kinematic amplification rates and saturation strengths at fixed resolution, appear attributable to varying levels of numerical diffusion. Notably, we find for this setup no systematic advantage of constrained transport over divergence-cleaning methods. We stress that this comparison, conducted in a highly idealised setting, represents a first step. Future extensions to more complex and physically realistic configurations remain essential.
\end{abstract}


\newcommand{\dotprod}[2]{{#1} \bm \cdot {#2}}
\newcommand{\avg}[1]{\left< {#1} \right>}
\newcommand{\kpara}{k_\parallel}
\newcommand{\kBcrossJ}{k_{\bm B \bm \times \bm J}}
\newcommand{\kBdotJ}{k_{\dotprod{\bm B}{\bm J}}}
\newcommand{\krms}{k_\mathrm{rms}}
\newcommand{\klambda}{k_\lambda}
\newcommand{\gradB}{\bm \nabla \bm B}
\newcommand{\bhat}{\hat{\bm b}}
\newcommand{\divB}{\bm \nabla \bm\cdot \bm B}

\section{Introduction}

Many astrophysical systems are highly turbulent and will quickly amplify any small magnetic seed field exponentially via a small-scale dynamo \citep{Brandenburg2005,Brandenburg2012}. However, numerical simulations of astrophysical objects, for example of the interstellar medium, a galaxy, or a galaxy cluster, can usually only model a small fraction of the involved huge range of scales \citep{Vogelsberger2020,Crain2023,Feldmann2026}. Therefore, simulations of galaxies in their full cosmological environment cannot resolve both the viscous and resistive scales. Thus in these astrophysical simulations, the smallest growth scale of the fluctuating small-scale dynamo is typically set by numerical dissipation, which is orders of magnitudes larger than the physical dissipation in the interstellar medium and in galaxies. In other words, the numerical resolution is limited to scales many orders of magnitude larger than the scales on which physical viscosity or resistivity would act. 

This is a significant complication for modelling the small-scale dynamo, because it grows the magnetic field fastest on the smallest scales. It is therefore directly sensitive to the details of the numerical method that set the diffusivity on these scales. Thus, the behaviour and sometimes outcome of the small-scale dynamo depend directly on the numerical choices made in the magnetohydrodynamics (MHD) solver, with consequences for any simulation results that potentially depend on magnetic fields.

In particular, cosmological simulations of galaxy formation are extremely demanding because of the large range of scales involved and the variety of barely resolved or fully unresolved physical processes happening at the smallest scales of the interstellar medium. Therefore, it is not surprising that different numerical codes have seen small-scale dynamos with markedly different properties in this setup \citep{pakmor2013,Rieder2016,Pakmor2017,Rieder2017,Rieder2017b,MartinAlvarez2018,Pfrommer2022,Wissing2023,Pakmor2024}. For MHD, these differences are exacerbated by the variety in techniques used. There are, in particular, codes that formally keep the divergence of the magnetic field exactly zero for some chosen discretization \citep{Evans1988}, and techniques that tolerate non-vanishing divergence provided it is kept small enough \citep[for example,][]{powell1999,dedner2002}.
Past comparison studies have found that codes implementing fairly different schemes can broadly agree, in particular in the case of decaying turbulence \citep{kritsuk_comparing_2011}.
For more idealized test problems, \citet{Tomida2026} recently investigated the impact of the divergence control technique on numerical results, and found important differences in some cases.
A comparison study is still missing for the more complex case of driven turbulence, which is closer to the physical conditions encountered in galaxy formation and in many other astrophysical systems. In parallel, numerical simulations have looked in detail into the small-scale dynamo in high resolution turbulent box simulations with explicit viscosity and resistivity. These simulations resolve all scales in the flow and directly control the Reynolds and Prandtl numbers, so they can assess their impact on the amplification rate and saturation level of the magnetic field \citep{Schekochihin2004,Kriel2022,Beattie2023}.

Here we compare different general purpose astrophysical MHD codes in a simple setup of driven subsonic turbulence in a periodic box. We focus on relatively low numerical resolution and purely numerical viscosity and resistivity, because this is the regime that essentially all galaxy formation simulations operate in. Slightly subsonic turbulence is also the typical regime for the diffuse warm phase and the volume filling hot phase of the interstellar medium \citep{Elmegreen2004} and the circumgalactic medium of galaxies \citep{Tumlinson2017}, the potentially most important places for a small-scale dynamo to operate in during galaxy formation. We aim to isolate the differences among the numerical schemes, in particular between schemes that conserve a divergence free magnetic field and schemes that allow for finite divergence, and the effect of their differences on the properties of the small-scale dynamo. To this end we implement an exactly identical turbulent driving in all codes, and run them through an identical analysis pipeline.

We first summarise the codes we employ and their configurations in Section~\ref{sec:codes}, describe the setup of our simulations and the shared analysis pipeline in Section~\ref{sec:setup}, and present a detailed description of our turbulent driving in Section~\ref{sec:driving}. We then discuss and compare the amplification of the magnetic fields and their saturation in Section~\ref{sec:amplification}. We analyse structural properties of the magnetic fields in the saturated regime in Section~\ref{sec:structures}, and present magnetic and kinetic power spectra of our simulations in the kinematic and saturated regimes in Section~\ref{sec:powerspectra}. We put our result into context and discuss the failure of the Arepo-CT scheme in Section~\ref{sec:discussion}. Finally, we end with a summary and outlook in Section~\ref{sec:summary}.

\section{Methods}

We first provide a brief overview of the different codes and their individual configuration choices we use. We then describe in detail the common setup of our simulations and their analysis. Throughout the rest of this paper, we use Gaussian units for the magnetic field,
where the magnetic energy density is $\bm B^2/(8\uppi)$.

\subsection{Codes}
\label{sec:codes}

\paragraph*{\Arepo} is a finite-volume code on an unstructured Voronoi mesh \citep{springel2010,pakmor2016,weinberger2020}. The mesh-generating points determine the Voronoi mesh and move with the gas velocity of each cell with an additional velocity correction to keep the mesh regular. Here we employ the standard choice in \Arepo, where the mesh-generating points of sufficiently distorted cells move towards their geometric centre with the sound speed of the cell \citep{Vogelsberger2012}. The second order finite volume scheme is described in detail in \citet{pakmor2016}. We use the Harten-Lax-van-Leer-discontinuity (HLLD) Riemann solver \citep{miyoshi2005} to compute fluxes over interfaces \citep{pakmor2011}. Here we disable refinement and de-refinement, so the number of cells is kept fixed. For the highest resolution  {\Arepo} simulations with $256^3$ cells they reach masses as low as $0.25\times$ and as high as $2.5\times$ the original cell mass. Thus, the standard Lagrangian mass refinement of {\Arepo} would refine and de-refine a small number of cells. We will leave simulations with explicit refinement and de-refinement to future study. 

We use global time steps only, that is all cells are evolved on the currently smallest time step of any cell in the simulation. Note that the Courant criterion in {\Arepo} does not need to take the gas velocity into account as the fluxes are computed in a co-moving frame, so the time step of a cell $\Delta t_\mathrm{cell}$ is computed as
\begin{equation}
  \Delta t_\mathrm{cell} = C_\mathrm{cfl} \frac{r_\mathrm{cell}}{c_\mathrm{f,cell}},
\end{equation}
where $r_\mathrm{cell}$ is the radius of the cell that we compute from the cell volume $V_\mathrm{cell}$ as $r_\mathrm{cell}=\left(3/(4\uppi)V_\mathrm{cell}\right)^{1/3}$, $c_\mathrm{f,cell}$ is the fast magneto-acoustic wave speed of the cell, and $C_\mathrm{cfl}=0.4$ is the Courant factor.

We use two different versions of the MHD solver that differ in their treatment of the divergence constraint $\divB=0$. The \emph{\ArepoPowell} version uses the Powell 8-wave scheme \citep{powell1999}. It adds source terms to advect the local divergence of the magnetic field $\divB/\rho$ with the mass flux, which quickly cancels most of the divergence that dominantly arises from the continuous change of the discretisation of the gradient operator in {\Arepo} \citep{pakmor2013}. The source terms restore Galilean invariance of the MHD equations, even when magnetic fields are not strictly divergence-free \citep{powell1999}. The implementation of the Powell scheme requires a choice on how to exactly discretise the source terms, but this does not add any parameters. We use a finite difference discretisation of the source terms described in detail in \citet{pakmor2013}. In particular, we calculate the divergence of the magnetic field in a cell from the Gauss integral of normal components of the magnetic field on the interfaces and take all other quantities in the Powell source terms as the cell centred values. Note that with the source terms our discrete equations do not inherently conserve total momentum any more to machine precision.

The \emph{\ArepoDedner} scheme uses the Dedner divergence cleaning method \citep{dedner2002} instead of the Powell scheme. It introduces an additional scalar $\Psi$ that absorbs the divergence of the magnetic field, and is actively transported away and dissipated. The Dedner scheme introduces two additional parameters, the speed with which $\Psi$ propagates ($c_\mathrm{h}$) and the timescale on which it dissipates ($\tau$). We use the implementation described in detail in \citet{pakmor2011}. In every time step we determine the propagation velocity $c_\mathrm{h}$ as the maximum of the fast magneto-acoustic wave speed of all cells. We then set the dissipation timescale to be $\tau=2r_\mathrm{cell}/c_\mathrm{h}$, generalising the choice from the original paper to universal problems.

Finally, we do not include the \emph{\ArepoCT}  scheme in the formulation of \citet{Mocz2016} in our comparison, because we now know that it is prone to numerical instabilities at high resolution, as we will discuss more in Section~\ref{sec:arepoct}. 

\begin{figure*}
	\centering
	\includegraphics[width=\textwidth]{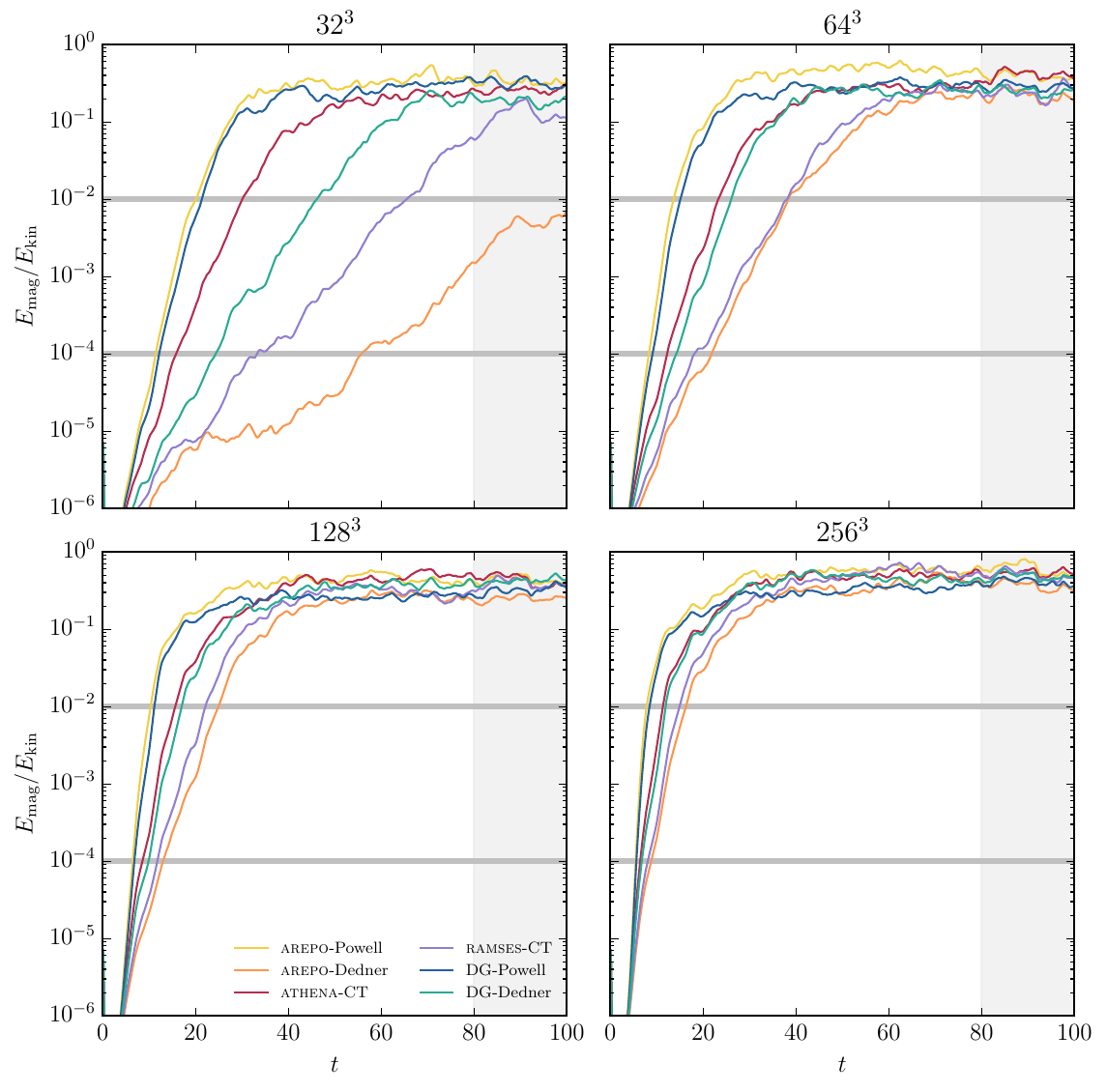}
	\caption{Amplification of magnetic fields over time. The panels show the time evolution of the ratio of magnetic to kinetic energy for all four resolution levels from $32^3$ (top left) to $256^3$ (bottom right). Each panel shows one line for each code. The gray horizontal lines show the threshold values between which the magnetic field amplification rate is fitted in the kinematic regime. The gray shaded area on the right of each panel shows the time interval over which we average properties to estimate them in the saturated state. All codes behave qualitatively similar, they show faster amplification and earlier saturation at higher resolution. The saturation level of the magnetic energy is similar for all codes.}
	\label{fig:amplificationAll}
\end{figure*}

\begin{table*}
	\begin{center}
		\setlength\tabcolsep{5pt}
        \begin{tabular}{ l r r r r r r r r }
        \hline
        Code \& scheme & Resolution & $\left<\mathcal{M}\right>$ & $\left<\frac{E_\mathrm{mag}}{E_\mathrm{kin}}\right>$ & $\Gamma$ & $t_0$ & $t_1$ & $N_\mathrm{cpu}$ & CPU hours\\
        \hline
        \ArepoPowell & $32^3$ & $0.17$ & $0.34$ & $0.53$ & $11.30$ & $  20$ & $8$ & $4.4\times 10^{1}$\\
        \ArepoPowell & $64^3$ & $0.18$ & $0.39$ & $0.89$ & $8.30$ & $  14$ & $112$ & $8.2\times 10^{2}$\\
        \ArepoPowell& $128^3$ & $0.18$ & $0.43$ & $1.18$ & $6.40$ & $  10$ & $896$ & $1.4\times 10^{4}$\\
        \ArepoPowell & $256^3$ & $0.17$ & $0.61$ & $2.00$ & $5.40$ & $   8$ & $2304$ & $2.0\times 10^{5}$\\
        \hline
        \ArepoDedner & $32^3$ & $0.22$ & $0.00$ & $-$ & $-$ & $-$ & $8$ & $4.5\times 10^{1}$\\
        \ArepoDedner & $64^3$ & $0.20$ & $0.22$ & $0.28$ & $21.80$ & $  38$ & $112$ & $8.7\times 10^{2}$\\
        \ArepoDedner & $128^3$ & $0.20$ & $0.25$ & $0.38$ & $12.90$ & $  25$ & $896$ & $1.4\times 10^{4}$\\
        \ArepoDedner & $256^3$ & $0.19$ & $0.37$ & $0.61$ & $8.80$ & $  16$ & $2304$ & $1.9\times 10^{5}$\\
        \hline
        \AthenaCT & $32^3$ & $0.20$ & $0.26$ & $0.32$ & $15.80$ & $  30$ & $64$ & $1.3\times 10^{1}$\\
        \AthenaCT & $64^3$ & $0.18$ & $0.41$ & $0.42$ & $12.20$ & $  23$ & $64$ & $3.3\times 10^{1}$\\
        \AthenaCT & $128^3$ & $0.18$ & $0.44$ & $0.66$ & $8.50$ & $  16$ & $64$ & $2.9\times 10^{2}$\\
        \AthenaCT & $256^3$ & $0.18$ & $0.53$ & $0.94$ & $6.40$ & $  11$ & $512$ & $8.9\times 10^{3}$\\
        \hline
        \RamsesCT & $32^3$ & $0.20$ & $0.13$ & $0.14$ & $33.20$ & $  65$ & $8$ & $1.5\times 10^{0}$\\
        \RamsesCT & $64^3$ & $0.19$ & $0.24$ & $0.23$ & $18.10$ & $  38$ & $112$ & $3.3\times 10^{1}$\\
        \RamsesCT & $128^3$ & $0.18$ & $0.39$ & $0.45$ & $11.70$ & $  22$ & $112$ & $4.9\times 10^{2}$\\
        \RamsesCT & $256^3$ & $0.17$ & $0.50$ & $0.68$ & $8.00$ & $  15$ & $1680$ & $9.6\times 10^{3}$\\
        \hline
        \DGPowell & $32^3$ & $0.19$ & $0.32$ & $0.50$ & $12.10$ & $  21$ & $8$ & $8.9\times 10^{0}$\\
        \DGPowell & $64^3$ & $0.20$ & $0.29$ & $0.77$ & $9.00$ & $  15$ & $64$ & $1.8\times 10^{2}$\\
        \DGPowell & $128^3$ & $0.20$ & $0.32$ & $1.05$ & $6.70$ & $  11$ & $256$ & $3.6\times 10^{3}$\\
        \DGPowell & $256^3$ & $0.19$ & $0.41$ & $1.59$ & $5.50$ & $   8$ & $2304$ & $1.6\times 10^{5}$\\
        \hline
        \DGDedner & $32^3$ & $0.20$ & $0.19$ & $0.21$ & $24.30$ & $  46$ & $8$ & $8.5\times 10^{0}$\\
        \DGDedner & $64^3$ & $0.20$ & $0.25$ & $0.39$ & $14.20$ & $  26$ & $32$ & $1.4\times 10^{2}$\\
        \DGDedner & $128^3$ & $0.18$ & $0.42$ & $0.64$ & $9.80$ & $  17$ & $112$ & $3.9\times 10^{3}$\\
        \DGDedner & $256^3$ & $0.18$ & $0.48$ & $0.90$ & $6.70$ & $  12$ & $2304$ & $1.7\times 10^{5}$\\
        \hline
        \end{tabular}
	\end{center}
	\caption{Summary of global properties of our turbulence simulations. The columns show, from left to right, (i) the code the simulation was run with, (ii) the number of resolution elements used for the simulation, (iii) the time-averaged Mach number $\left<\mathcal{M}\right>$, (iv) the ratio between magnetic and kinetic energy in the box $\left\langle E_\mathrm{mag}/E_\mathrm{kin}\right\rangle$ averaged between $t=80$ and $t=100$, (v) the growth rate of the magnetic field $\Gamma$ in the kinematic regime, estimated between (vi) the time $t_0$ when the magnetic energy first passes $10^{-4}$ of the kinetic energy and (vii)  $t_1$ when the magnetic energy first passes $10^{-2}$ of the kinetic energy (in code units), (viii) the number of compute cores we used for the simulation, and finally (ix) the core hours it took to run the simulation to $t=100$.}
	\label{tab:turbulence}
\end{table*}

\paragraph*{\ArepoDG,} also known as \Tenet, is a discontinuous Galerkin (DG) code on Cartesian adaptive mesh-refinement (AMR) grids based on the infrastructure of the {\Arepo} code. It relies on the DG schemes for hydrodynamics and MHD described in detail in \citet{schaal2015}, \citet{guillet2019}, \citet{cernetic2023}, and \citet{cernetic2024}. The scheme uses arbitrary-order Legendre polynomial expansions of the conserved variables in each of the cells, allowing for discontinuities at cell faces. The solution polynomial in each cell is evolved with both an internal contribution coming from the analytical flux function within the cell, and a contribution from the faces involving the solution of Riemann problems, for which the HLLD numerical flux \citep{miyoshi2005} is used. Integrals in the volume and faces of cells are computed using Gaussian quadrature. In this work, we only use a fixed Cartesian grid with adaptive mesh refinement disabled. We also limit ourselves to the second-order scheme where variables are expanded linearly within each cell. In this case, the cell averages and linear gradients in each cell are tracked and evolved self-consistently by the DG scheme. Therefore, the DG simulations have more degrees of freedom than the finite volume simulations. 

Note that unlike in finite volume codes, slopes are not reconstructed in DG. For the present subsonic turbulence test, we also do not use any oscillation limiter, but we have the positivity limiter described in \cite{guillet2019} enabled as a default. All runs of {\ArepoDG} in this work use the second-order SSP-RK2 integrator \citep{gottlieb2005} with Courant number $0.6$. 

For MHD, we applied the two different schemes studied in \cite{guillet2019} to control the divergence of the magnetic field. The \emph{\DGPowell} scheme uses a locally-divergence-free (LDF) vector basis to represent the magnetic field, ensuring $\divB = 0$ in the interior of each cell. At faces, the DG scheme does not impose continuity of the normal component of $\bm B$, therefore a local divergence will arise. In order to control it, {\DGPowell} uses a Powell-like source term \citep{powell1999} at the faces. In this work, the Powell term follows equations~(31) and (33) from \cite{guillet2019}.

The \emph{\DGDedner} scheme, unlike \DGPowell, does not use an LDF basis for the magnetic field, instead it expands each of the components of $\bm B$ independently with Legendre polynomials. As a result, the scheme will produce a finite $\divB$ both within the cell and at the faces from discontinuities of the normal $\bm B$ component. The scheme uses the divergence cleaning approach of \cite{dedner2002} to advect and damp $\divB$. We refer the reader to \cite{guillet2019} for a detailed description of the DG implementation. In this work, for {\DGDedner} runs, we set the hyperbolic cleaning speed $c_\mathrm{h}$ at every time step to be $1.05$ times the fastest signal speed in the box, and the damping parameter is set to $c_\mathrm{r}=0.18$ following the original prescription of \cite{dedner2002}. Note that this choice is in general not suitable for all problems because $c_\mathrm{r}$ is a dimensional quantity; it is however satisfactory for this turbulent box where time and length scales are close to unity.

\paragraph*{\Ramses} is a finite volume code on a Cartesian grid with a tree-based adaptive mesh refinement scheme \citep{teyssier2002,teyssier2006}. Here we use a fixed uniform Cartesian mesh, and disable adaptive mesh refinement. MHD is implemented via the staggered grid CT scheme \citep{Evans1988}. The magnetic field is stored on cell interfaces, where each interface evolves the component of the magnetic field normal to that interface. Cell-centred magnetic fields are reconstructed from the interface values when needed. 

The magnetic field is evolved such that the discrete cell-centred divergence of the magnetic field is zero by construction. Moreover, the normal component of the magnetic field is continuous at interfaces by construction, so the scheme is also manifestly divergence free for the one-dimensional (1D) Riemann problem solved at each interface. Interface fluxes are computed using the HLLD Riemann solver \citep{miyoshi2005}. The update of the magnetic fields requires computing the electro-motive force at the edges of the interfaces. {\Ramses} does this with a two-dimensional (2D) version of the HLLD Riemann solver similar to the method described in \cite{balsara_2012}. 

Since we use only a single refinement level with a uniform mesh, all cells evolve with a global time step, set by a Courant factor of $0.8$. {\Ramses} implements several options for slope limiters in both the interface flux calculation and the magnetic field update. We use the monotonised central limiter for both types of slope limiters for all simulations shown here. We use $\gamma=1.01$.

\begin{figure*}
	\centering
	\includegraphics[width=\textwidth]{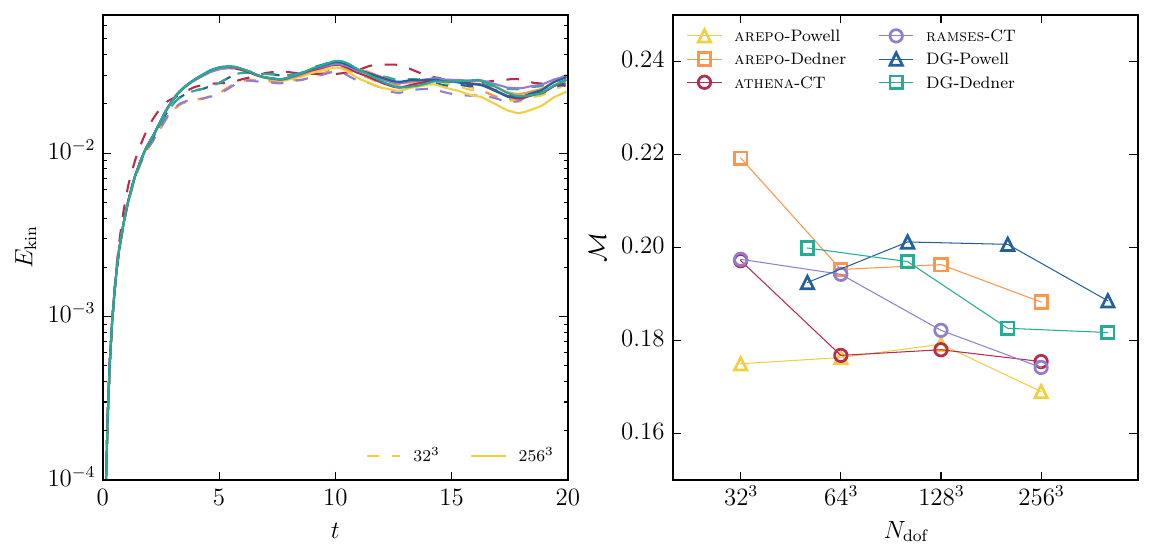}
	\caption{The left panel shows the time evolution of the kinetic energy in the box focusing on the early time of the simulation $t<20$ for all codes at a resolution of $32^3$ and $256^3$. All simulations have saturated the kinetic energy at $t=5$, and it remains roughly constant afterwards. The right panel shows the mean Mach number $\mathcal{M}$ for different resolutions and numerical schemes; see the text for the definition for $\mathcal{M}$. The values are time averaged between $t=80$ and $t=100$ where most simulations have saturated their magnetic dynamo. All schemes saturate at a very similar turbulent kinetic energy.}
	\label{fig:machnumber}
\end{figure*}

\paragraph*{\AthenaPP} is a finite volume code on a Cartesian grid with a block-based adaptive refinement strategy \citep{stone2008,stone2020}. We use it with a fixed uniform Cartesian mesh, without adaptive mesh refinement. Similarly to \RamsesCT, MHD is implemented via the staggered grid CT scheme \citep{Evans1988}. Note that there are minor differences in how the electro-motive force is computed between \AthenaCT and \RamsesCT. In particular, \AthenaCT uses an upwind average of the values from four cells touching an edge, rather than solving the full 2D Riemann problem as is done in {\Ramses}.

In our default setup for \AthenaPP, we use the piecewise parabolic method (PPM) \citep{Colella1984} and a third order Runge-Kutta time integrator with the HLLD Riemann solver \citep{miyoshi2005}. Moreover,  we apply the slope limiter in characteristic variables, rather than primitive variables as for all other codes we employ. This configuration aims to reduce the numerical diffusivity, and therefore the numerical viscosity and numerical resistivity as much as possible. We show a comparison of difference choices for the setup in \AthenaPP (using a slope limiter in primitive variables or going down to second order linear reconstruction and a second order time integrator) in Section~\ref{sec:discussion}. We run \AthenaPP in isothermal mode with $\gamma=1.01$, and with a Courant factor of $0.3$. 

\subsection{Setup and analysis}
\label{sec:setup}

Our simulations use a cubic box of size $L=1$ with periodic boundary conditions. Initially it is uniformly filled with an ideal gas at rest with constant density, $\rho=1$, and constant pressure, $p=1$. We initialise the simulation with a uniform magnetic seed field of strength $B_z=10^{-4}$ along the $z$-direction of the box. This corresponds to an initial magnetic energy density of $4\times10^{-10}$. 

All simulations, including the {\Arepo} moving mesh simulations, start with a uniform Cartesian grid. We run all codes at four different resolution levels from the lowest resolution of $32^3$ to the highest resolution of $256^3$. Note that we use the same number of cells also for the \ArepoDG\ simulations, so those have a slightly larger number of degrees of freedom (by a factor $4^{1/3}$ at fixed number of cells), because the cell centred gradients are stored and evolved as well.

We aim to compare the different simulations at the same nominal resolution. We still anticipate differences because of different effective numerical viscosity and resistivity in each simulation. We expect that all codes reduce the numerical viscosity and resistivity with increasing numerical resolution at a similar rate, and therefore increase the amplification rate of the turbulent dynamo in the kinematic regime at a similar rate. Finally, we will compare simulations of different codes at different numerical resolution to understand if their results agree when they seemingly have similar effective numerical viscosity or resistivity.

We simulate an approximately isothermal equation of state. We set the adiabatic constant of the gas to $\gamma = 1.01$, and reset the pressure of the gas in every cell $i$ at every time step so that the local sound speed in the cell is unity,
\begin{equation}
	P_i \leftarrow c_\mathrm{s}^2 \rho_i / \gamma, \quad c_\mathrm{s} = 1,
	\label{eq:reset-P}
\end{equation}
similar to the technique used in \citet{bauer_subsonic_2012}.

We evolve all simulations until $t=100$. For the {\Arepo} and {\ArepoDG} simulations we run the analysis on the fly to compute and output various properties of the simulation with a cadence of $\Delta t=0.1$. We also output $9$ full snapshots between $t=60$ and $t=100$ with a cadence of $\Delta t=5$.

To avoid having to implement the same on-the-fly postprocessing in {\Athena} and {\Ramses} as well, we instead produce $1000$ full snapshots for all {\Athena} and {\Ramses} simulations with a cadence of $\Delta t=0.1$. We then convert their format into {\Arepo}-readable snapshots,  fully conserving the mesh structure and the cell-centred gas properties. This allows us to apply exactly the same analysis code in postprocessing as done on-the-fly in the {\Arepo} runs, and to compute the same properties for our further comparison.

For our analysis of the DG schemes, we use the dynamically evolved slopes to obtain cell-centred gradients. For all other finite volume codes, including {\Athena} and {\Ramses}, we compute cell-centred gradients as done in  {\Arepo} \citep{pakmor2016}. As a result, the cell-centred gradient estimates used internally by {\Athena} and {\Ramses} can differ slightly from those computed in our analysis, even when applied to the same mesh and fluid quantities. This is particularly important for the gradient tensor of the magnetic field: our computation of the gradient will \emph{not} guarantee that the divergence of the magnetic field is zero on {\Athena} and {\Ramses} provided solutions, because of the slightly different definitions adopted for the  discrete gradient operators.

\begin{figure}
	\centering
	\includegraphics[width=\columnwidth]{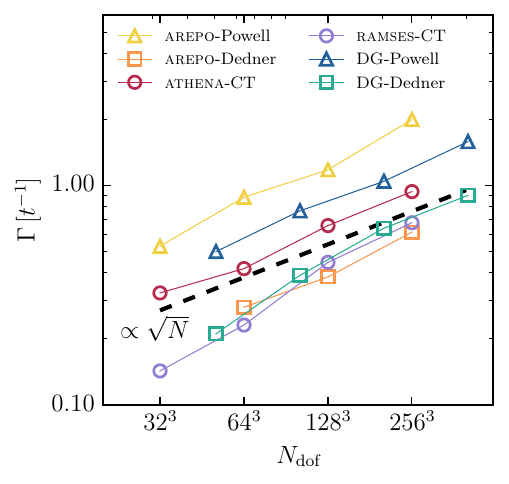}
	\caption{Exponential amplification rate of the magnetic energy in the \emph{kinematic} regime for different resolutions and numerical schemes. The rate is measured  between the times when the ratio of magnetic energy to kinetic energy in the box first reaches $10^{-4}$ and $10^{-2}$ (see horizontal lines in Figure~\ref{fig:amplificationAll}). All schemes seem consistent with an increase of the amplification rate with resolution that is slightly steeper than $\propto \sqrt{N}$ but differ in their normalisation because of their different numerical viscosity and hence, effective Reynolds numbers. The amplification rate of {\ArepoDedner} at a resolution of $32^3$ is not shown because it is below the scale (see also Table~\ref{tab:turbulence}).}
	\label{fig:amplification}
\end{figure}

\begin{figure}
	\centering
	\includegraphics[width=\columnwidth]{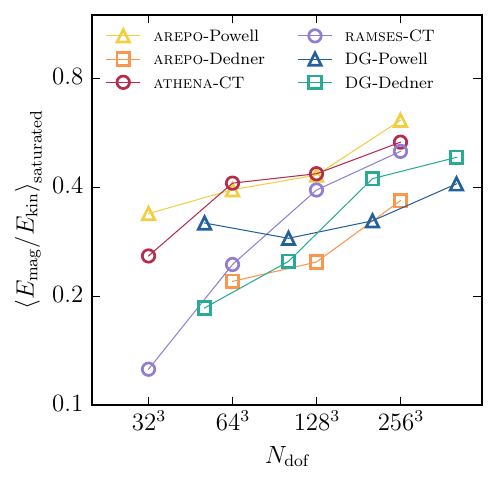}
	\caption{Time averaged saturation of the magnetic energy for different resolutions and numerical schemes. We compute this quantity as the time average between $t=80$ and $t=100$ of the magnetic-to-kinetic energy ratio. Note that the $x$-axis is logarithmic. All schemes reach saturation values within a factor of two for a resolution of $64^3$ or better. The saturation strength slightly increases with better resolution in all schemes. The saturation strength of {\ArepoDedner} at a resolution of $32^3$ is not shown because it has not yet reached saturation at the end of the simulation (see also the top right panel of Figure~\ref{fig:amplificationAll}).}
	\label{fig:saturation}
\end{figure}

\subsection{Turbulent driving}
\label{sec:driving}

To drive turbulence in the box, we inject momentum via a stochastic force density field $\bm f(\bm x, t) = \rho \bm a(\bm x, t)$, which we have implemented in an identical way for all codes. Driving turbulence with a force density field rather than an acceleration field $\bm a$ guarantees the conservation of total momentum in the simulation. The driving follows an established general methodology \citep[see, for example][]{schmidtNumericalDissipationBottleneck2006,federrath_comparingstatisticsinterstellar_2010,bauer_subsonic_2012}.
The force density $\bm f(\bm x, t)$ is expressed as a superposition of Fourier modes $\bm k$ that are periodic in the box, with time-dependent complex amplitudes:
\begin{equation}
	\bm f(\bm x, t) = \sum_{\bm k} \frac{2}{3} \operatorname{Re}\left[ \bm{A}(\bm k, t) e^{i\bm k \bm\cdot \bm x} \right].
	\label{eq:drive-f}
\end{equation}
For each mode $\bm k$, the complex amplitude vector $\bm{A}(\bm k, t) \in \mathbb{C}^3$ is determined from a power spectrum $P$ and a complex stochastic noise vector $\bm{N}(\bm k, t) \in \mathbb{C}^3$:
\begin{equation}
	\bm{A}(\bm k, t) = \sqrt{P(k)} \left( \bm{N}(\bm k, t) - \frac{1}{k^2}i \bm k \bm\cdot \bm{N} (\bm k, t) \right).
	\label{eq:drive-A_m}
\end{equation}
This expression ensures that $\bm f$ is purely solenoidal (divergence-free).

For $P(k)$, we follow \cite{federrath_comparingstatisticsinterstellar_2010} and use a parabolic amplitude spectrum that injects power for wavenumbers $k \in [k_\text{min}, k_\text{max}]$, with peak power $P_0$:
\begin{align}
	\sqrt{P(k)}  & = \sqrt{P_0} \cdot \max \left(0,  1-\frac{4(k - k_\text{mid})^2}{(k_\text{max} - k_\text{min})^2} \right), \\
	k_\text{mid} & \equiv \frac{k_\text{min} + k_\text{max}}{2}.
\end{align}
We set $k_\text{min} = 2\uppi/L$ and $k_\text{max} = 3 \times 2\uppi/L$, so that the peak injection scale $k_\text{mid}$ corresponds to a length scale $L/2$. The driving amplitude $\sqrt{P_0}$ is set to $2\times 10^{-4}$, which yields an overall subsonic Mach number $\mathcal{M} \approx 0.2$ in steady-state.

\begin{figure*}
	\centering
	\includegraphics[width=\textwidth]{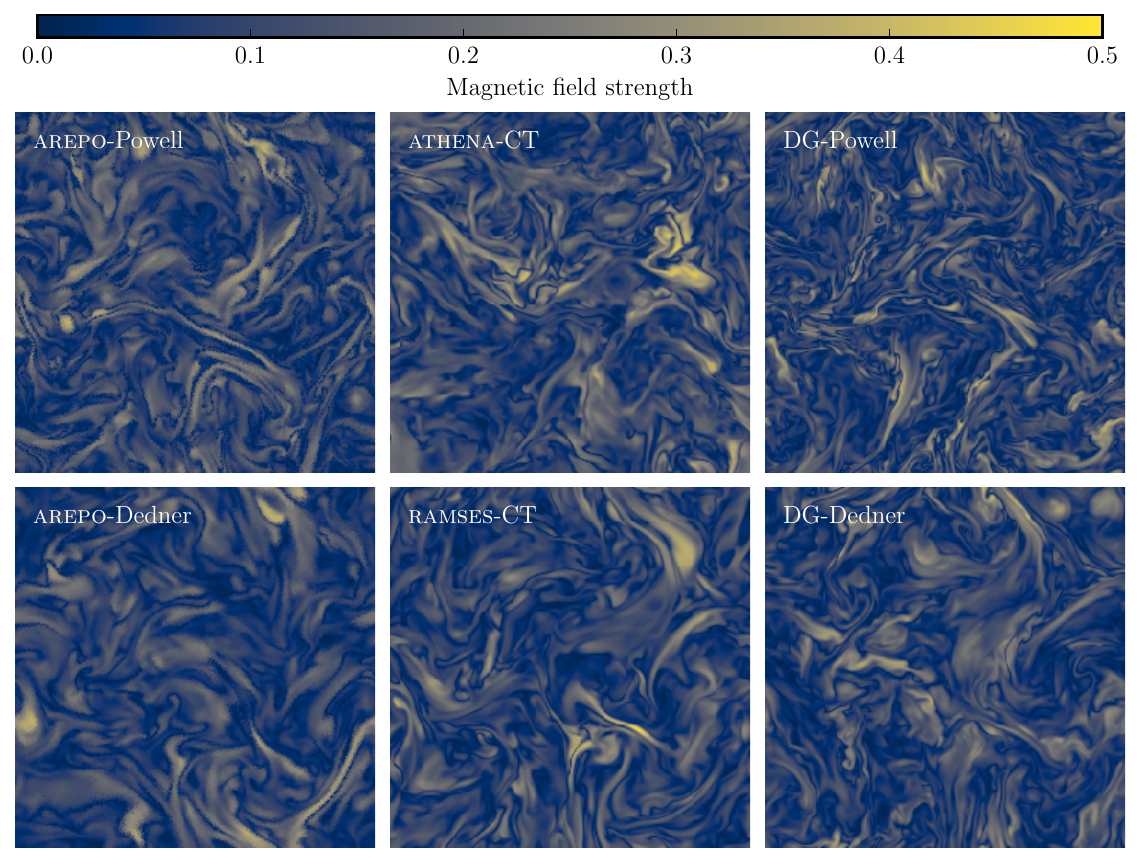}
	\caption{Slices of magnetic field strength at $t=100$ for all six schemes at the highest numerical resolution of $256^3$. The overall field strength and large scale structures are similar, but the amount of small scale structure varies significantly. {\DGPowell} shows the smallest structures, owing to its additional degrees of freedom. For {\DGDedner}, these structures are likely smeared out by the Dedner divergence cleaning scheme. The {\ArepoDedner} and {\RamsesCT} simulations show least small scale structure.}
	\label{fig:bfld_slices}
\end{figure*}

The stochastic noise vector $\bm{N}(\bm k, t)$ is constructed from 6 independent realizations of a real-valued Ornstein-Uhlenbeck process $X(t)$ (real and imaginary parts for 3 vector components). This random process is advanced forward in time $X_n \to X_{n+1}$ at regular time intervals of $\Delta t = 0.005$ time units, with:
\begin{equation}
	X_{n+1} \leftarrow \theta X_n + \sigma \sqrt{1-\theta^2} N_n, \quad \theta \equiv e^{-\Delta t/\tau},
\end{equation}
where the $N_n \sim \mathcal{N}(0, 1)$ are independent normally-distributed random numbers. We set the process variance $\sigma^2$ and autocorrelation time $\tau$ to $\sigma^2=1$ and $\tau = 1$, respectively. Updating the process with a fixed $\Delta t$ disconnected from the hydrodynamical times is important to ensure reproducibility of the random driving across resolutions and codes. We use the exact same driving with the same realisation of random phases for all simulations of all codes. The range of scales we drive turbulence on is fully resolved even for the $32^3$ simulations. So the driving is also exactly the same across any resolution levels. 

For {\Ramses}, {\Athena}, {\Arepo} and {\ArepoCT}, the force is computed at the centre of each cell.
For {\ArepoDG}, the force is evaluated at all internal quadrature points within each cell.

\section{Amplification and saturation of magnetic fields}
\label{sec:amplification}

We first show an overview of the amplification and saturation process of the magnetic energy relative to the kinetic energy in Figure~\ref{fig:amplificationAll}. The magnetic energy for all simulations starts with an initial phase of exponential amplification. Except for some of the most diffusive schemes at $32^3$ (see upper left panel of Figure~\ref{fig:amplificationAll}), all simulations eventually reach saturation for the magnetic energy, which then remains roughly constant until the end of the simulation. For all simulations with a resolution of $64^3$ or higher the saturation strength of the different schemes agrees within a factor of~$2$. For all simulation codes, the growth rate increases with resolution during the initial exponential amplification phase. We will quantify the amplification rates and their change with resolution in more detail below.

We show the time evolution of the kinetic energy for the $32^3$ and $256^3$ simulations at early times ($t<20$) and the average Mach number for all simulations in Figure~\ref{fig:machnumber}. The kinetic energy for all simulations already saturates before $t=5$ and then stays roughly constant. The average Mach number is computed over the timespan from $t=80$ to $t=100$ (shaded gray area in Figure~\ref{fig:amplificationAll}).  We compute the Mach number for one snapshot of one simulation as
\begin{equation}
\mathcal{M}=\sqrt{\frac{2\,E_\mathrm{kin}}{M\,c_\mathrm{s}^2}},
\end{equation}
where $E_\mathrm{kin} = \sum_i \frac{1}{2}m_i\,\bm{\varv}_i^2$ is the total kinetic energy in the box computed from the velocity $\bm{\varv}_i$ and mass $m_i$ of all cells, $M=1$ is the total mass in the box, and $c_\mathrm{s}=1$ is the isothermal sound speed. All simulations have time-averaged Mach numbers close to $0.2$, which slightly decrease with higher resolution due to the increase in magnetic energy.

We show the amplification rate of the magnetic field in the kinematic regime for all simulations in Figure~\ref{fig:amplification}. To measure the amplification rate, we compute an exponential timescale from the times when the ratio of magnetic energy to kinetic energy in the box first reaches $10^{-4}$ and when it first reaches $10^{-2}$ as
\begin{equation}
\Gamma = \frac{ \log{\frac{10^{-2}}{10^{-4}}} }{ t_{10^{-2}} - t_{10^{-4}} }.
\end{equation}
As seen qualitatively already in Figure~\ref{fig:amplificationAll}, the amplification rate increases with resolution for all simulation codes with a similar slope, independently of the details of the numerical scheme and its implementation. This indicates that the change of numerical dissipation with resolution is the critical factor that determines the amplification rate. The amplification rate scales with resolution slightly steeper than $\propto \sqrt{N}$, with different normalisations for each scheme.

Both dissipation processes---numerical viscosity that dissipates kinetic energy, and numerical resistivity that dissipates magnetic energy---occur on the grid scale and therefore move to smaller scales with higher resolution. The different numerical implementations will typically yield different numerical viscosity and resistivity.

The dependence of the amplification rate and the saturation field strength of the turbulent dynamo for subsonic turbulence is not fully understood. High resolution numerical simulation with explicit viscosity and resistivity which have fixed kinetic and magnetic Reynolds numbers indicate that the amplification rate in the kinematic regime dominantly depends on the magnetic Reynolds number \citep{Kriel2022}. So we expect it to depend mainly on the numerical resistivity in our simulations. The saturation strength of the magnetic field rather seems to depend mostly on the Prandtl number $\mathrm{Pm}=\mathrm{Rm}/\mathrm{Re}$, the ratio of the magnetic Reynolds number $\mathrm{Rm}$ to the kinetic Reynolds number $\mathrm{Re}$ \citep{Kriel2022}. 

For simulations that share the same hydrodynamic solver but use different magnetic-field schemes, we find that the differences in dynamo growth rates are consistent with the expectation that the kinematic growth rate depends primarily on numerical resistivity. Comparing the {\ArepoPowell} with the {\ArepoDedner} simulations and the {\DGPowell} with the {\DGDedner} simulations, we see that the Powell simulations feature significantly faster amplification rates compared to the Dedner simulations, despite both pairs using identical hydrodynamics solver in a regime where the magnetic field is dynamically irrelevant. This difference is slightly larger for \Arepo\ than for DG.

We also see that the linear reconstruction CT schemes and Dedner schemes yield very similar amplification rates across codes. This is likely coincidence, however, as changing for example the Riemann solver or the slope limiter of either simulation significantly affects the amplification rate. The Powell schemes feature an amplification rate up to 3--4$\times$ higher at fixed resolution compared to the linear reconstruction CT scheme {\RamsesCT} and up to a factor of two higher compared to the higher order PPM scheme that we use for {\AthenaCT}.

In Figure~\ref{fig:saturation}, we show the ratio of magnetic to kinetic energy for all simulations, again time averaged between $t=80$ to $t=100$. The saturation strength increases with resolution for all simulations codes. The differences in saturation strength between the different codes at fixed resolution seem to decrease slightly with higher resolution. 

For a visual impression of the structure of the magnetic field in the saturated regime, we show slices of the magnetic field strength at $t=100$ for all simulation codes at a resolution of $256^3$ in Figure~\ref{fig:bfld_slices}. The overall strength and large-scale structures look similar. The amount of small scale structure and the smallest scales on which structures exist vary significantly, however. In particular, the {\DGPowell} simulation appears to resolve smaller scales than the other simulations, which is likely a combination of the low diffusivity of the Powell scheme, the reduced hydrodynamical diffusivity of DG compared to finite-volume schemes (because slopes are retained at every timestep), and the additional degrees of freedom available in the DG scheme. The \ArepoPowell simulation seems to be slightly noisier on the smallest scales than the other simulations.

\begin{figure*}
	\centering
	\includegraphics[width=\textwidth]{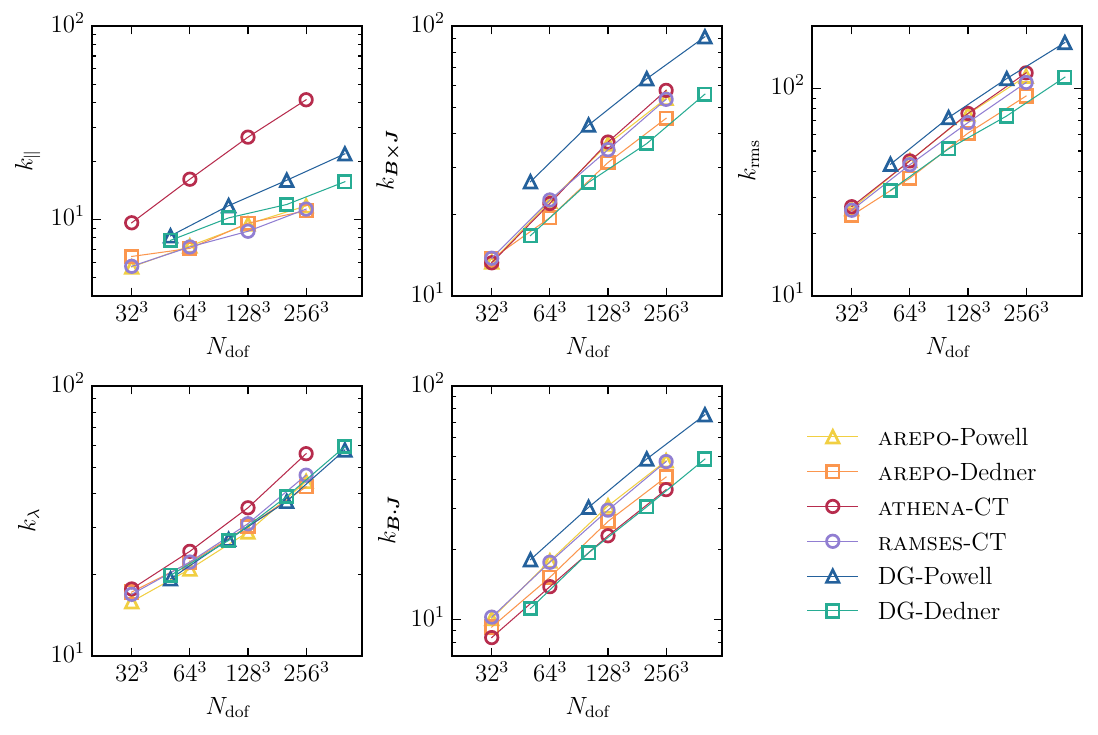}
	\caption{
		Various global measures of structural properties of the magnetic field in the \emph{saturated} regime for different resolutions and numerical schemes. We show time averages of the quantities between $t=80$ and $t=100$. The different schemes span a range of values at fixed numerical resolution, without any code or scheme standing out in any obvious way for any of the metrics. All measures increase with resolution for all schemes, because the viscous scale decreases for all schemes at higher resolution. There is no clear systematic difference between schemes that enforce $\divB = 0$ by construction and schemes that employ divergence control.
        }
	\label{fig:kstats}
\end{figure*}

\begin{figure*}
	\centering
	\includegraphics[width=\textwidth]{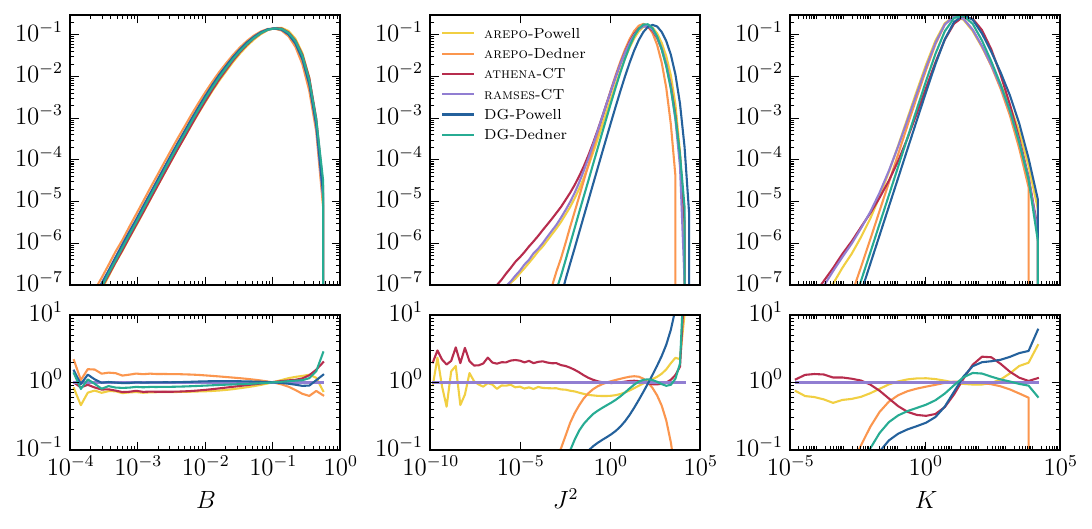}
	\caption{Time averaged volume-weighted histograms of the $\log_{10}$ of the magnetic field strength $B$, current squared $J^2$, and curvature $K$ of the \emph{saturated} magnetic field for the different numerical schemes at a numerical resolution of $256^3$. We stack histograms between $t=80$ and $t=100$. The top row shows the histograms, the bottom panels the histograms relative to {\RamsesCT}. All distributions are very similar. Notably {\AthenaCT}, {\RamsesCT}, and {\ArepoPowell} are able to represent significantly smaller values of $J^2$ and $K$.
    The DG runs have a distribution of $J^2$ and $K^2$ that is significantly shifted to larger values, possibly because we used the dynamically evolved slopes to compute $K$ and $J$, which are updated by the DG scheme without any limiter.
    }
	\label{fig:hist1d}
\end{figure*}

\begin{figure*}
	\centering
	\includegraphics[width=\textwidth]{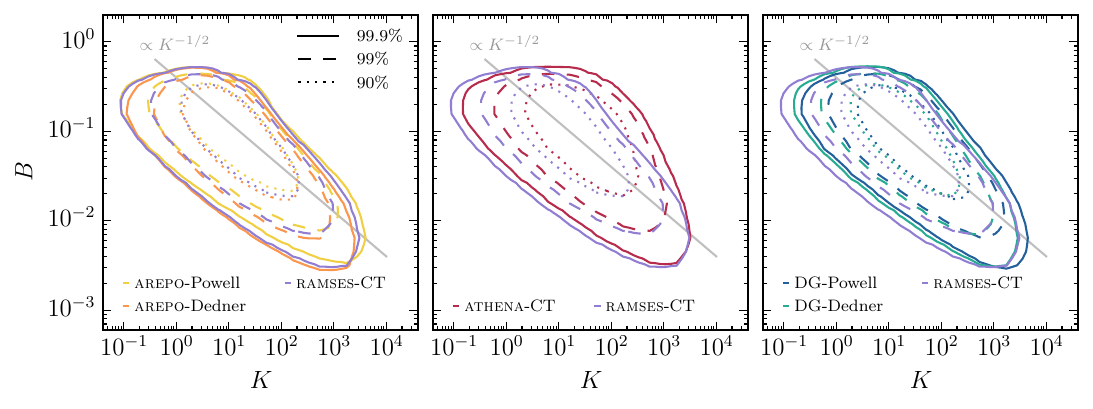}
	\caption{Contours of the time averaged volume weighted 2D histogram of magnetic field strength $B$ versus curvature $K$ in the \emph{saturated} regime. The histograms are for the different numerical schemes at a numerical resolution of $256^3$ and stacked between $t=80$ and $t=100$. We include the histogram of {\RamsesCT} in all panels as a reference to guide the eye. All distributions are overall similar and consistent with the expected slope of $\propto K^{-1/2}$ \citep{Schekochihin2004} indicated by the gray line. Again, the {\ArepoDG} distributions are slightly shifted to higher values of $K$ by a factor of about 2.
    {\AthenaCT} stands out with a broader distribution in the upper right quadrant (large $B$, large $K$), showing convex contours, unlike all other codes which present a slight concave ``dip''.
    }
	\label{fig:hist2d}
\end{figure*}

\section{Structural properties of the saturated magnetic field}
\label{sec:structures}

Following \cite{Schekochihin2004}, we define characteristic local wave numbers of the magnetic field as:
\begin{align}
	\kpara    & = \left( \frac{\avg{ \left| (\dotprod{\bm B}{\bm\nabla}) \bm{B} \right|^2 }}{\avg{B^4}} \right)^{1/2}, \label{eq:k-para}   \\
	\kBcrossJ & = \left( \frac{\avg{ \left| {\bm B} \bm \times {\bm J} \right|^2 }}{\avg{B^4}} \right)^{1/2}, \label{eq:k-BcrossJ} \\
	\kBdotJ   & = \left( \frac{\avg{ \left| \dotprod{\bm B}{\bm J} \right|^2 }}{\avg{B^4}} \right)^{1/2}, \label{eq:k-BdotJ}   \\
	\krms     & = 
    \left[\frac{1}{\langle\bm{B}^2\rangle/2}
    \int_0^\infty \mathrm{d}k k^2 \epsilon_\mathrm{mag}(k)\right]^{1/2}=
    \left( \frac{\avg{ \left| \gradB \right|^2 }}{\avg{B^2}} \right)^{1/2}. \label{eq:k-rms}
\end{align}
where $\bm J = \bm \nabla \bm \times \bm B$ is the MHD current, $\avg{ \cdot }$ denotes volume-weighted averaging over the whole domain, and $\epsilon_\mathrm{mag}(k)=\frac{1}{2}\int\mathrm{d}\Omega_{\bm{k}} k^2\langle|\bm{B}(\bm{k})|^2\rangle$ is the angle-integrated magnetic energy spectrum.
$\kpara$ is the wave number along the local field direction, $\kBcrossJ$ is the wave number in the perpendicular direction, $\kBdotJ$ is the wave number in the direction orthogonal to both $\bm B$ and $\bm B \bm \times \bm J$, and $\krms$ is the overall local root mean squared wave number of the gradient length of the magnetic field.
Similarly, we also compute the root mean squared wave number for the gradient length of the velocity field:
\begin{align}
	\klambda & = 
    \left[\frac{1}{\langle\bm{u}^2\rangle/2}
    \int_0^\infty \mathrm{d}k k^2 \epsilon_\mathrm{kin}(k)\right]^{1/2}=
    \left( \frac{\avg{ \left| \bm \nabla \bm u \right|^2 }}{\avg{u^2}} \right)^{1/2}, \label{eq:k-lambda}
\end{align}
where $\epsilon_\mathrm{kin}(k)$ is the angle-integrated kinetic energy spectrum.
We also compute the magnetic curvature $K$:
\begin{equation}
	K = \left| \left(\bhat\bm\cdot\bm \nabla\right) \bhat \right|, \quad \bhat = \bm B / B.
	\label{eq:K}
\end{equation}

These quantities involve expressions of the form $\avg{f(\partial_i B_j)}$ where $f$ is a nonlinear function, and are computed differently for {\ArepoDG} and the finite volume codes. In the DG code, the $\partial_i B_j$ are readily obtained at any point inside any cell using the polynomial expansion of the $B_j$ variables, without any reconstruction. We can therefore evaluate $f(\partial_i B_j)$ at the DG quadrature points, and obtain the average over each cell (and thereafter over the whole domain) from the quadrature rule.
For all the finite volume codes (including {\Ramses} and {\Athena}), we first use {\Arepo}'s slope-limited reconstruction algorithm to obtain reconstructed gradients $\widetilde{\partial_i B_j}$ at the centre of each cell. We then approximate the average over each cell to second-order as $\avg{f(\partial_i B_j)} \approx f(\widetilde{\partial_i B_j})$.

Finally, the wave number values shown in Figure~\ref{fig:kstats} are the time averaged mean values between $t=80$ and $t=100$. We first note that there are no codes that are global outliers in all or even most plots. All measures increase with resolution for all simulation codes with a similar slope for the respective definitions of wave number, implying that they are all determined by numerical dissipation. All {\Arepo} simulations are very close to the {\RamsesCT} simulations in all measures shown in Figure~\ref{fig:kstats}. The differences between Powell and Dedner do not seem to affect the global structural properties of the magnetic fields in the saturated state shown here.

The {\AthenaCT} simulations are similar to {\Arepo} and {\RamsesCT} for most measures, but feature significantly larger $\kpara$ and somewhat smaller $\kBdotJ$. Both codes implement CT schemes, but with notable differences in the reconstruction order, calculation of the electro-motive force, and details of the limiting procedure. Those differences are likely responsible for these discrepancies.

For a more detailed comparison of the structural properties of the magnetic fields in the saturated regime, we show time averaged volume weighted histograms of magnetic field strength, current, and curvature of the magnetic field for the $256^3$ resolution simulations in Figure~\ref{fig:hist1d}. The histograms are stacks of $200$ equally spaced outputs between $t=80$ and $t=100$.

The magnetic field histograms are very similar between all simulation codes. The differences are limited to small shifts from values above $B=10^{-1}$ to values below. The histograms of {\AthenaCT} and {\ArepoPowell} are essentially identical. 

The histograms of the current are overall similar but show some interesting differences. The current values for the {\DGPowell} schemes are systematically shifted to slightly larger values by a factor of $\approx 2$. The {\ArepoDedner} simulation notably does not reach the largest $J^2$ values. Moreover, the CT simulations as well as the {\ArepoPowell} simulation show a tail to smaller currents that is absent in the DG simulations and the {\ArepoDedner} simulation. The closest agreement here is between {\RamsesCT} and {\ArepoPowell}.

The curvature histograms again look overall similar. The {\ArepoDedner} scheme sticks out because of the unique absence of cells with the highest curvature values that are present in all other simulations. Similar to the current, the CT simulations and {\ArepoPowell} show a tail to small curvature values that is absent in the other simulations. This may indicate that they are able to better retain strong and ordered or laminar fields. The closest agreement here is again between {\RamsesCT} and {\ArepoPowell}.

Finally, we show two-dimensional histograms of the magnetic field strength and curvature in Figure~\ref{fig:hist2d}. We expect both quantities to be physically related, roughly scaling as $B\propto K^{-1/2}$ (indicated for reference as solid gray lines in the Figure, see \citealt{Schekochihin2004}). We also show the contours of {\RamsesCT} in the background of the left and right panels for comparison. All simulations roughly follow the expected scaling. As expected from the 1D histograms (see Figure~\ref{fig:hist1d}), for both DG simulations the 2D histogram is slightly shifted to higher curvature, but without changing the slope of the relation or the shape of the 2D histogram. The 2D histograms of the {\Arepo} simulations are both almost identical to the histogram of the {\RamsesCT} simulation. Interestingly {\RamsesCT} and {\AthenaCT} differ noticeably at large magnetic field strengths and curvatures. There, {\AthenaCT} uniquely features a tail to high field strengths at large curvature that is absent in all other simulations.

\begin{figure}
	\centering
	\includegraphics[width=\columnwidth]{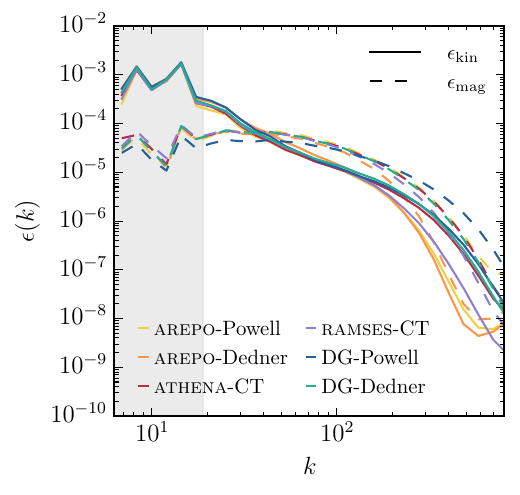}
	\caption{Kinetic and magnetic power spectra in the \emph{saturated} regime of the magnetic field for the different numerical schemes at a numerical resolution of $256^3$, time-averaged between $t=80$ and $t=100$. The shaded grey area on the left indicates the scales on which we drive turbulence. The kinetic power spectra are essentially identical on large scales, but differ slightly on small scales where numerical dissipation sets the shape. The magnetic power spectra share the same shape and normalisation at large scales, with the notable exception of the {\DGPowell} scheme that has a slightly lower normalisation. On small scales, where numerical dissipation sets the shape of the magnetic power spectra, they differ slightly between the different schemes.}
	\label{fig:powerspectra}
\end{figure}

\begin{figure}
	\centering
	\includegraphics[width=\columnwidth]{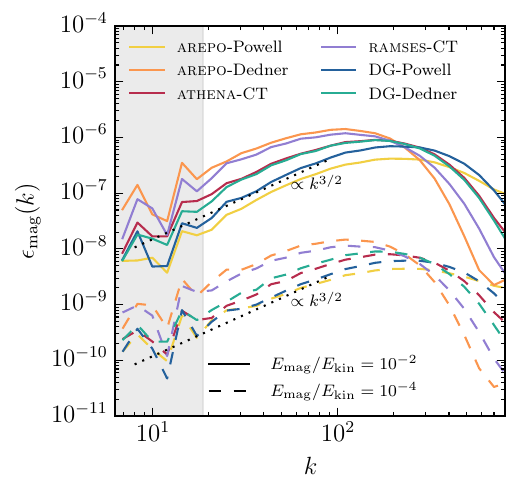}
	\caption{Magnetic power spectra for different schemes in the \emph{kinematic} regime at a resolution of $256^3$. We show two simulation times for each scheme, when the total magnetic energy first surpasses $10^{-4}$ and $10^{-2}$ of the total kinetic energy, respectively. The shaded grey area indicates the scales on which we drive turbulence.  The shapes of the power spectra on large scales are all consistent with the Kazantsev slope \citep{kazantsev1985}. By construction, the normalisation on large scales at these times is larger for schemes that have less small scale power, owing to higher numerical dissipation  (because the integrals of the power spectra are identical across the codes).}
	\label{fig:powerspectra_amplification}
\end{figure}

\begin{figure}
	\centering
	\includegraphics[width=\columnwidth]{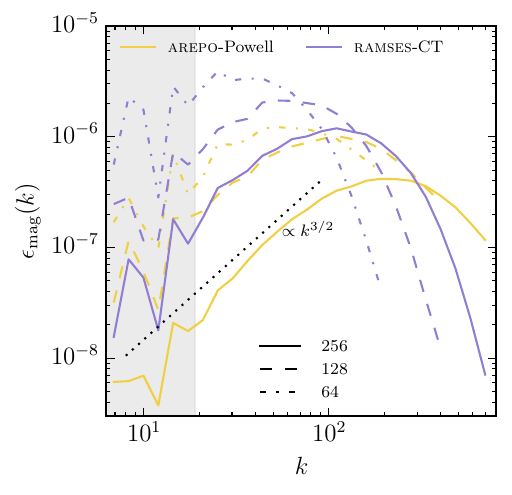}
	\caption{Magnetic power spectra for {\RamsesCT} and {\ArepoPowell} in the \emph{kinematic} regime at the time when the total magnetic energy first surpasses $10^{-2}$ of the total kinetic energy for different numerical resolutions. The shaded grey area indicates the scales on which we drive turbulence. The shape of the power spectra on large scales are all consistent with the Kazantsev slope \citep{kazantsev1985}. On small scales the power spectra of both schemes evolve in similar ways with resolution but their shape differs. The power spectrum of {\ArepoPowell} at a resolution of $128^3$ is very similar to the power spectrum of {\RamsesCT} at a resolution of $256^3$.}
	\label{fig:powerspectra_amplification_resolution}
\end{figure}

\begin{figure}
	\centering
	\includegraphics[width=\columnwidth]{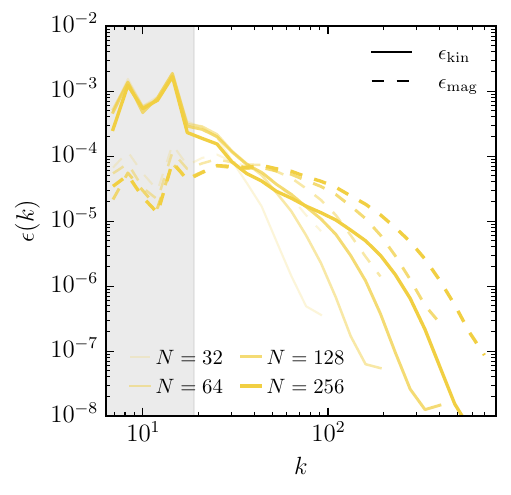}
	\caption{Magnetic and kinetic power spectra for {\ArepoPowell} in the \emph{saturated} state for different numerical resolutions. The shaded grey area indicates the scales on which we drive turbulence. On large scales the power spectra are nearly converged even for the lowest resolution shown. With higher resolution they extend more faithfully to smaller scales. }
	\label{fig:powerspectra_resolution_arepo}
\end{figure}

\section{Magnetic and kinetic power spectra}
\label{sec:powerspectra}

To understand how kinetic and magnetic energy are distributed on different scales for the different simulations and how this changes with the amplification of the magnetic field we look at power spectra of the kinetic and magnetic energy.

To compute power spectra, we first remap the vector field of interest to a high-resolution $512^3$ uniform grid. For the finite volume codes, the field values are interpolated to the centres of the cells of the power spectrum grid, using gradients as reconstructed by the {\Arepo} code. For {\ArepoDG}, the field values are evaluated directly at the centre of each cell of the power spectrum grid using the DG polynomials.

For the kinetic energy power spectrum we Fourier transform the quantity $\sqrt{\frac{1}{2}\rho}\,\bm{\varv}$ and for the magnetic energy power spectrum we use $\sqrt{\frac{1}{8\uppi}}\,\bm{B}$. Thus, the integral over each power spectrum yields the total energy of its component.

In Figure~\ref{fig:powerspectra}, we first look at kinetic and magnetic power spectra for the highest resolution $256^3$ simulations in the saturated regime, averaged over $200$ equally spaced outputs from $t=80$ to $t=100$. The kinetic power spectra are essentially identical on large scales, as expected because we drive turbulence only on large scales with the exact same modes in all simulations. They differ on small scales though, most likely due to a result of different numerical dissipation for the different numerical schemes and implementations. Notably, {\AthenaCT} and the DG simulations have more power on small scales in the kinetic energy, consistent with lower viscosity from a slightly larger number of degrees of freedom in the DG simulations and the higher order reconstruction scheme used in {\AthenaCT}. The {\Arepo} simulations and {\RamsesCT} are very similar and only differ on the smallest scales. There the kinetic power spectrum becomes flat for the {\Arepo} simulations, likely a consequence of the broader spectrum of dissipation scales originating from the spatially varying geometry of individual cells around the Nyquist scale.

The shape of the magnetic power spectrum on large scales in the saturated state is very similar again. Notably the {\DGPowell} simulation has less power on large scales than the other simulations, but it has the largest power on small scales. The other schemes have consistent magnetic power spectra up to $k\approx 70$. On smaller scales, that is for larger $k$, the spectra differ in the wave number at which the power begins to roll off. They are ordered in small scale power in the same way as the total saturated energy shown in Figure~\ref{fig:saturation}. In other words, the difference in total magnetic energy in the saturated state originates fully from the smallest scales for all but the {\DGPowell} simulation.

In Figure~\ref{fig:powerspectra_amplification} we show magnetic power spectra during the kinematic phase of the turbulent dynamo when the magnetic field is irrelevant for the dynamics. We show magnetic power spectra for the highest resolution $256^3$ simulations at the time when the total magnetic energy in each simulation first reaches $10^{-4}$ (dashed lines) and $10^{-2}$ (solid lines) of the total kinetic energy of the simulation at this time. On large scales the slope of the power spectra of all simulations is consistent with $\epsilon_\mathrm{magnetic}\propto k^{3/2}$, as expected for a turbulent dynamo \citep{kazantsev1985}.

Because the total kinetic energy in all boxes is roughly constant with time, the magnetic power spectra shown here also represent the same total magnetic energy, even though they are taken at different times (see Figure~\ref{fig:amplificationAll}). Therefore, simulations with more power on small scales will have a lower normalisation on large scales. On small scales the slope, and in particular the scale where the magnetic power turns over is different for all simulations.

To understand how numerical resolution affects the power spectra in the saturated regime, in Figure~\ref{fig:powerspectra_amplification_resolution} we show magnetic power spectra for {\RamsesCT} and {\ArepoPowell} as an example for resolution levels from $64^3$ to $256^3$ at the time when the total magnetic energy first surpasses $10^{-2}$ of the total kinetic energy in each simulation. All magnetic power spectra are consistent with the Kazantsev slope \citep{kazantsev1985} on large scales. For both codes the power spectra gain more power on small scales with higher resolution, which leads to a lower normalisation on larger scales at higher resolution. The shift to smaller scales is comparable between the two codes. Notably the $256^3$ magnetic power spectrum of {\RamsesCT} is almost identical to the $128^3$ power spectrum of {\ArepoPowell}.

Finally, in Figure~\ref{fig:powerspectra_resolution_arepo} we show time-averaged kinetic and magnetic power spectra in the saturated regime between $t=80$ and $t=100$ exemplary for {\ArepoPowell} for all resolution levels from $32^3$ to $256^3$. On the largest scales just below the injection scale, the kinetic and magnetic power spectra are nearly independent of numerical resolution. The scale at which the kinetic power spectra deviates from the highest-resolution reference shifts to larger scales as resolution decreases. The difference in scales is very similar for each factor of two in resolution, again suggesting that the difference is purely driven by numerical dissipation, and the result becomes more faithful with higher numerical resolution. The normalisation of the kinetic power spectrum on the scale just below the injection scales decreases at the highest resolution, likely because the magnetic field starts to react back as its saturation strength increases (see Figure~\ref{fig:saturation}). Still, the shape of the power spectra for scales larger than the deviation point are converged. The magnetic power spectra seem to be fully converged from scales just below the injection scales to the point where they deviate from the next better resolution simulation.

\begin{figure*}
	\centering
	\includegraphics[width=\textwidth]{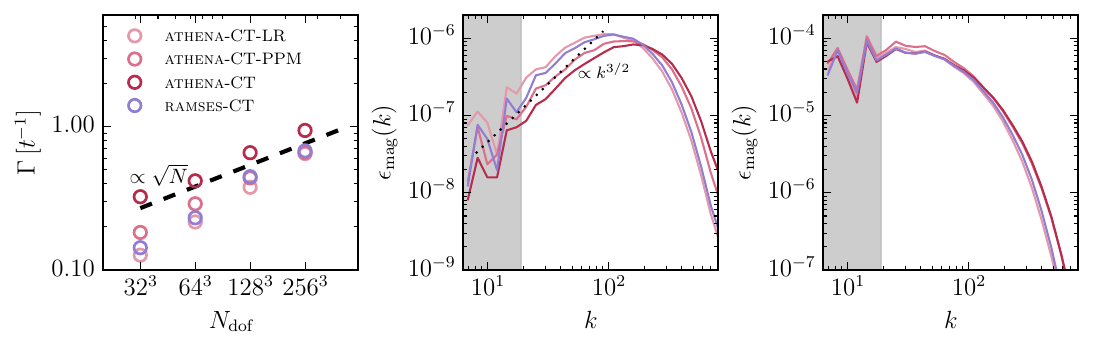}
	\caption{
		Comparison of {\Athena} with different settings for reconstruction and slope limiter, including a second order scheme with linear reconstruction (LR), a higher order version using third order time integration and the piecewise parabolic method for reconstruction (PPM), and our default choice with third order time integration, the piecewise parabolic method for reconstruction, and applying the slope limiter in characteristic variables ({\AthenaCT}) rather than in primitive variables. We also show {\RamsesCT} which uses a second order linear reconstruction scheme with slope limiting in primitive variables (similar to LR). The panels show the amplification rate of the magnetic field between the times when the magnetic energy reaches $10^{-4}$ and $10^{-2}$ of the kinetic energy for different numerical resolution (left panel), the magnetic power spectra in the \emph{kinematic} regime at the time when the magnetic energy reaches $10^{-2}$ of the kinetic energy at a resolution of $256^3$ (middle panel), and the magnetic power spectra averaged between $t=80$ and $t=100$ in the \emph{saturated} regime also at a resolution of $256^3$ (right panel). Using higher order and  characteristic limiters leads to faster amplification of the magnetic field. At the highest resolution, using characteristic-variable instead of primitive-variable limiters has a larger effect than going from linear reconstruction to the piecewise parabolic method. In the \emph{saturated} regime the difference in the magnetic power spectra between the PPM simulations with different slope limiters almost disappears.
        }
	\label{fig:grid_comparison}
\end{figure*}

\section{Discussion}
\label{sec:discussion}

\subsection{The influence of numerical diffusivity}

All simulations we compare here implement grid-based discretisation schemes, but they differ significantly in their hydrodynamics and MHD solvers. None of them explicitly include viscosity and resistivity. Instead, both are set by numerical diffusivity on the grid scale.

Despite fundamental differences between the simulations, in particular in the treatment of the divergence constraint of MHD, there is no indication for any of the simulations to be a systematic outlier in the properties we study for our idealised periodic boxes of driven subsonic turbulence with a weak seed magnetic field. The differences we observe are compatible with differences in numerical dissipation. Schemes with lower resistivity have faster amplification rates of the magnetic field in the kinematic regime \citep{Kriel2022}, and more magnetic energy on smaller scales at the same total magnetic energy (see Figure~\ref{fig:powerspectra_amplification}). Notably, \ArepoPowell at $128^3$ resolution and \RamsesCT at $256^3$ resolution have almost identical magnetic power spectra at the time when each simulation first reaches a ratio of magnetic to kinetic energy of $10^{-2}$ (see Figure~\ref{fig:powerspectra_amplification_resolution}). At this time they are still well in the kinematic regime.

The magnetic energy in the saturated regime increases with resolution for all schemes, and does so roughly by the same amount for the same increase in resolution (see Figure~\ref{fig:saturation}). This seems to be dominantly caused by the presence of additional magnetic energy on smaller scales when they become resolved (see Figure~\ref{fig:powerspectra_resolution_arepo} that shows this as an example for \ArepoPowell). A similar effect has been observed for simulations with controlled viscosity and resistivity when the Prandtl number (ratio between magnetic and kinetic Reynolds number) increases \citep{Brandenburg2011, Kriel2022, Beattie2023}.

It is not completely straightforward, however, to connect this to the increase we observe. In our simulations with numerical viscosity and resistivity, increasing the resolution will always reduce the scale of dissipation and increase the magnetic and kinetic Reynolds numbers by a roughly similar amount. There are indications that for grid codes (which at least use a non-moving mesh) both Reynolds numbers do not increase exactly at the same rate with resolution, but the magnetic Reynolds number increases slightly more than the kinetic Reynolds number \citep{grete_2023}. This leads to a small but systematic increase of the Prandtl number with increasing resolution, that could contribute to the increase in the saturated magnetic field strength \citep{Kriel2022}. Note also that dissipation is more complicated in moving mesh codes than in fixed grid codes, because the different cell sizes (that also locally vary with time) spread the dissipation over a range of scales rather than fixing it to a specific scale as in  Cartesian fixed-grid codes (see Figure~\ref{fig:powerspectra}).

\subsection{The influence of different MHD solvers}

Despite the differences in schemes, we can learn from comparing subsets of our simulations that share some properties between their hydrodynamics and MHD solvers. When we compare codes that use the same hydrodynamics scheme but different MHD solvers (that is comparing {\ArepoDedner} with {\ArepoPowell} and {\DGDedner} with {\DGPowell}) in the kinematic regime, we can directly see the differences caused by the different MHD solvers. In this regime the magnetic fields are dynamically irrelevant, so the effective viscosity of the simulations with the same hydrodynamics solver are exactly the same, yet we observe different amplification rates of the magnetic field strength. {\ArepoPowell} has an amplification rate that is $3\times$ faster than {\ArepoDedner}, and {\DGPowell} amplifies the magnetic field about $2\times$ faster than {\DGDedner}. We interpret these differences as a consequence of the different MHD implementations that lead to a different effective resistivity. Moreover, the different ratios for {\Arepo} and DG show that it is not only the implementation of the magnetic fields, but also the combination with the mesh and hydrodynamics scheme that matters for the amplification rate of the magnetic field.

\begin{figure}
	\centering
	\includegraphics[width=\columnwidth]{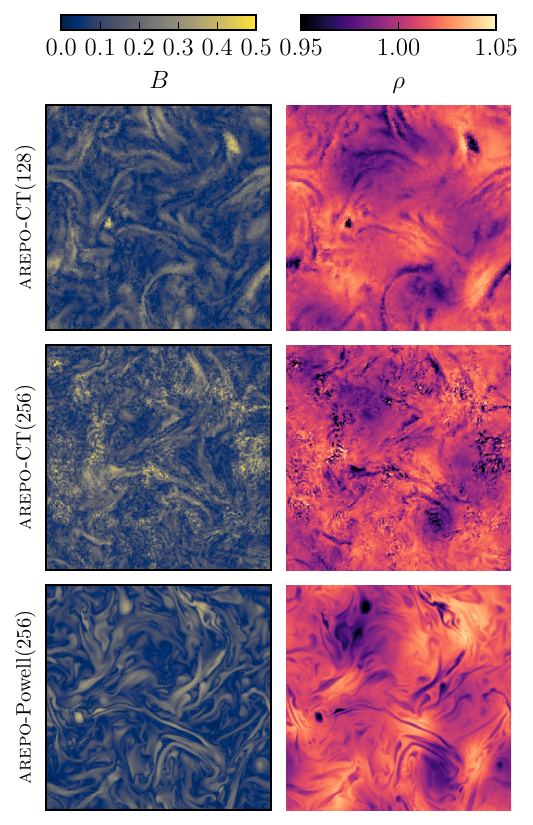}
	\caption{Slices of the magnetic field strength at $t=100$ (left column) and density (right column) for {\ArepoPowell} (at a resolution of $256^3$) and {\ArepoCT} at resolutions of $128^3$ and $256^3$. At $256^3$ the {\ArepoCT} scheme becomes obviously unstable and shows prominent oscillatory fluctuations on small scales. A hint of these instabilities is already visible in the slices at a resolution of $128^3$, but they are still subdominant.}
	\label{fig:slices_arepo_ct}
\end{figure}

\begin{figure*}
	\centering
	\includegraphics[width=\textwidth]{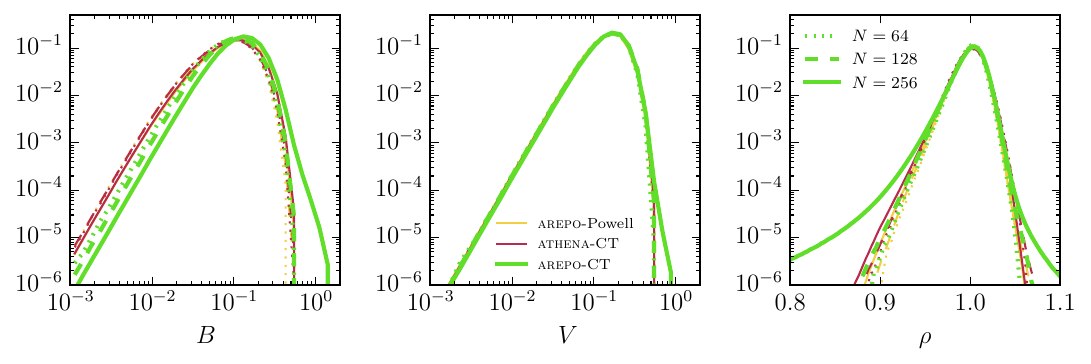}
	\caption{Histograms of the $\log_{10}$ of the magnetic field strength $B$, velocity magnitude $V$, and density $\rho$ for {\ArepoCT} at different numerical resolutions, similar to Figure~\ref{fig:hist1d}. The histograms are stacked from $t=80$ to $t=100$. We show the same histograms for {\ArepoPowell} and {\AthenaCT} in the background for reference. {\ArepoCT} at $256^3$ resolution is a clear outlier, again showing an excess of large fluctuations in the magnetic field, the velocity field, and most obviously in the density.}
	\label{fig:hist1d_arepo_ct}
\end{figure*}

Our default configurations for {\AthenaCT} and {\RamsesCT} differ in the order of the scheme. {\RamsesCT} uses linear reconstruction with a slope limiter in primitive variables and a second order time integrator. For {\AthenaCT} we use an approximately third order PPM scheme with a third order time integrator, and a slope limiter in characteristic variables. These configurations are typical for the applications of both codes. {\Athena} is often used for rather well resolved, more idealised problems, that are well suited for higher order methods. {\Ramses}, by contrast, is primarily employed in cosmological simulations of galaxy formation, which present a fundamentally different numerical regime. The inherent physical complexity and turbulent multi-scale nature of such cosmological simulations combined with their typically limited spatial resolution frequently compromises the stability of higher-order methods, making lower-order schemes the more robust and practical choice.

To understand the impact of a higher order reconstruction scheme (PPM) compared to linear reconstruction, and of limiting in characteristic variables compared to limiting in primitive variables, we compare different setups of {\Athena} with {\RamsesCT} in Figure~\ref{fig:grid_comparison}. In particular, we compare the amplification rate in the kinematic regime as well as magnetic power spectra in the kinematic regime and the saturated regime. In addition to our default {\AthenaCT} and {\RamsesCT} simulations we add two additional sets of simulations with {\Athena}. {\AthenaCTLR} uses linear reconstruction, second order time integration, and limits in primitive variables similar to {\RamsesCT}, while {\AthenaCTPPM} uses PPM and limits in primitive variables as well.

In the left panel of Figure~\ref{fig:grid_comparison} we see that the magnetic amplification rates for {\AthenaCTLR} are very similar to {\RamsesCT}. The remaining differences are plausibly explained by differences in the calculation of the electromotive force. Limiting in characteristic variables significantly increases the amplification rate. At low resolution changing from linear reconstruction to PPM also increases it, but at higher resolution this difference eventually disappears. These differences add up so that {\AthenaCT} simulations have a significantly faster amplification rate in the kinematic regime than the {\RamsesCT} simulations (more than $2\times$ faster at $64^3$ resolution and still $40\%$ faster at $256^3$ resolution).

The magnetic power spectra in the kinematic regime shown in the middle panel of Figure~\ref{fig:grid_comparison} confirm that in this phase increasing the order from linear reconstruction to PPM and changing from limiting in primitive to characteristic variables both decrease the resistivity of the scheme and lead to more power on small scales. In the saturated regime, the magnetic power spectra of {\RamsesCT} and {\AthenaCTLR} as well as the power spectra of {\AthenaCTPPM} and {\AthenaCT} are essentially identical, showing that in this regime the order of the scheme makes a difference, but details of the limiting procedure less so.

There is in particular no systematic difference between the formally divergence free schemes (CT) and the divergence control and cleaning schemes (Powell and Dedner). The differences in the kinematic regime are consistent with different levels of dissipation that lead to different amounts of numerical resistivity at a given resolution. There are some differences between the different schemes in the structural properties of the magnetic field in the saturated regime (for various global structural measures in Figure~\ref{fig:kstats} and for distributions of magnetic field strength, current, and curvature of the magnetic field in Figures~\ref{fig:hist1d} and \ref{fig:hist2d}). However, the differences we see do not separate schemes by divergence treatment. Rather, other choices in the details of the numerical schemes seem to matter more here, something that could be an interesting target to explore in more detail in the future.

\subsection{Numerical instability of \ArepoCT}
\label{sec:arepoct}

To demonstrate that \ArepoCT in the implementation of \citet{Mocz2016} is unstable at high resolution in our turbulence problem, we show slices of the magnetic field strength and density at $t=100$ for the \ArepoCT simulations at resolutions of $128^3$ and $256^3$ in Figure~\ref{fig:slices_arepo_ct}, and compare those to slices of the \ArepoPowell simulation at $256^3$. Already at a resolution of $128^3$ the magnetic field strength and density slices of \ArepoCT show small-scale granularity that is absent in the \ArepoPowell simulation (or any other scheme, see Figure~\ref{fig:bfld_slices}). At a resolution of $256^3$ this becomes even more obvious. Clear signatures of local instability are evident that lead to artifacts in both the density and the magnetic field strength.

To quantify these artifacts we show 1D histograms of the magnetic field strength, velocity magnitude, and density stacked between $t=80$ and $t=100$ for \ArepoCT with the same histograms for \ArepoPowell and \AthenaCT in the background, for comparison. The histograms for \ArepoCT are still seemingly consistent with the other codes for $64^3$ and $128^3$, even though we can already see visual artifacts in the magnetic field strength and density slices for the $128^3$ simulation. For the $256^3$ \ArepoCT simulation, we clearly see the instabilities manifest as a tail to higher magnetic field strength, velocity, and density that is absent in any of the other simulations.

The particular formulation used in \ArepoCT for the induction equation is based on the idea of evolving a cell-averaged version of the vector potential,
\begin{equation}
\vec{Q}_i =\int_{V_i} \vec{A} \,{\rm d}V.
\end{equation}
This quantity is evolved in the Weyl gauge for every cell according to
\begin{equation}
\frac{{\rm d}\vec{Q}_i}{{\rm d}t}
= \int_{V_i} \vec{u} \bm\times \vec{B} \, {\rm d}V +  \int_{\partial V_i} \vec{A} \,\vec{w}\bm\cdot {\rm d} \vec{n},
\label{eqnarepoct}
\end{equation}
where $\vec{w}$ is the velocity with which the outer boundary of the cell $i$ moves, and ${\rm d}\vec{n}$ is an outwardly pointing differential area element. The first term on the right hand side is a source term due to the (negative) electric field, and the second term describes advection stemming from the mesh motion. To get from the cell-averaged vector potential to the magnetic field, \ArepoCT does not directly compute $\vec{B}=\bm\nabla \bm\times \vec{A}$ but rather constructs the magnetic flux through each face of the Delaunay tessellation, the topological dual of the Voronoi mesh, by means of a contour integral of the vector potential along the edges of a Delaunay face, $\Phi_B = \oint \vec{A}\bm\cdot {\rm d}\vec{l} $. This guarantees that the magnetic field is divergence-free by construction for all Delaunay tetrahedra as the sum of these magnetic fluxes will always add up to zero. From the magnetic fluxes, a volume averaged mean magnetic field is estimated for every Delaunay tetrahedron, which in turn is spatially averaged by calculating the overlaps with Voronoi cells to obtain estimates of the central magnetic fields of the Voronoi cells. 

Unfortunately, despite this elegant geometric construction and the fact that satisfactory results for various MHD test problems are obtained \citep[see][]{Mocz2016}, the scheme encounters numerical instabilities in complicated 3D flow as we have seen above. We think they are sourced to different degrees by several factors. One basic problem is that the vector potential is not really a conservative quantity. In fact, one can argue that the non-trivial part of the evolution of $\vec{Q}$ is entirely coming from the source term, that is the electric field, but that this term can easily become dominated by larger contributions from the advection term. Numerically, the spatial averaging of $\vec{A}$ over a cell, and the subsequent curl operation to obtain the $\vec{B}$-field (which is fundamentally what the integral reconstruction over faces achieves) do not commute, meaning that geometric effects from the local cell geometry can directly introduce local errors in the $\vec{B}$-field. This is further amplified by the fact that the Delaunay tessellation for a given distribution of mesh-generating points can change discontinuously based on an infinitesimal motion of one of the mesh-generating points -- this happens when one of the tetrahedra flips \citep[see][]{springel2010}. As a result, the estimated $\vec{B}$-fields at Voronoi cell centres can change discontinuously as well. This is already a potential source of numerical instability. A related problem concerns the fact that the vector potential is associated  with the location of the mesh generating point, whereas the spatial averaging over a cell would suggest that the vector potential $\vec{A}_i = \vec{Q}_i / V_i$, recomputed at the end of a time step from an updated value of $\vec{Q}_i$, should rather be associated with the geometric centre of a cell.

Another problem lies in the treatment of upwinding. With respect to the genuine advection term in $\vec{A}$, this is dealt with in \ArepoCT with a simple upwind prescription at each cell face. While this cures the violent instability that a simple forward-in-time, centred-in-space treatment of the advection term would create, the source term due to the electric field is simply evaluated in a cell-centred fashion, without any form of explicit upwinding. This is problematic, as it is known that this can produce  numerical instabilities in CT for the stationary mesh. For this reason, the fixed-grid CT codes use the Riemann solver solutions at cell interfaces to construct the electromagnetic forces via suitable averaging procedures.

Another conceptual weakness of \ArepoCT is that the formulation is not manifestly Galilean-invariant despite the use of a moving mesh. This means that a constant velocity boost for the whole simulation box is not simply causing translational motion of all primitive fluid quantities, including the magnetic field, but rather results in more work for the advection term in equation~(\ref{eqnarepoct}), and invariably a higher numerical diffusivity. Notably, this also means that the time step should be reduced accordingly for large bulk velocities, something that is however neglected by \ArepoCT in practice. Overcoming all of these numerical problems and to realize a generally stable, Galilean-invariant, moving-mesh formulation of constrained transport is thus still an elusive problem at this point. We have, however, made substantial progress on this goal recently (Springel et al., 2026, in prep.).

\section{Summary and outlook}
\label{sec:summary}

We presented a set of driven subsonic turbulent MHD box simulations of six different setups with four different astrophysical MHD codes and at four different levels of numerical resolution from $32^3$ to $256^3$. Starting with a small uniform magnetic seed field we analysed the properties of the small-scale turbulent dynamo first during the kinematic phase where the magnetic field grows exponentially and then during the saturated phase where the magnetic field strength remains roughly constant.

We compared global properties including the amplification rate of the magnetic field and its saturation level in Section~\ref{sec:amplification}. Most importantly we found significant differences in the amplification rate at fixed resolution. {\ArepoPowell} amplifies about $2\times$ faster than {\AthenaCT} and {\DGPowell}, which in turn amplify about $2\times$ faster than {\ArepoDedner}, {\RamsesCT}, and {\DGDedner}. Despite the differences in the value of the amplification rate at fixed resolution, all codes show a similar increase of the amplification rate with improved resolution close to $\propto \sqrt{N}$ as expected from the decrease of numerical viscosity and resistivity with resolution for a second-order scheme (see Figure~\ref{fig:amplification}).

In Section~\ref{sec:structures} we compared global structural properties of the magnetic fields in the saturated regime in Figure~\ref{fig:kstats}, histograms of magnetic field strength, current, and the curvature of the magnetic field in Figure~\ref{fig:hist1d}, and 2D histograms showing the correlation between magnetic field strength and the curvature of the magnetic field in Figure~\ref{fig:hist2d} and found that they are consistent between the different codes. Most critically, we did not find any differences that separate the codes according to the schemes that enforce the divergence constraint exactly ({\AthenaCT} and {\RamsesCT}) and the other methods that allow for a finite divergence combined with a cleaning suppression.

In Section~\ref{sec:powerspectra}, we compared power spectra of the magnetic field in both the kinematic regime (Figure~\ref{fig:powerspectra_amplification} and Figure~\ref{fig:powerspectra_amplification_resolution}) and the saturated regime (Figure~\ref{fig:powerspectra} and Figure~\ref{fig:powerspectra_resolution_arepo}). We again found differences in the kinematic regime that, however, seem to be consistent with effects expected from different levels of numerical diffusivity. Notably, we show in Figure~\ref{fig:powerspectra_amplification_resolution} that the magnetic power spectrum of {\RamsesCT} at a resolution of $256^3$ is almost identical to the magnetic power spectrum of {\ArepoPowell} at a resolution of $128^3$ at the time when each simulation first surpasses a ratio of magnetic to kinetic energy of $10^{-2}$.

We conclude that any of the six schemes of our comparison produces reasonable and consistent results for our very idealised setup of a small-scale dynamo in driven subsonic turbulence. Lastly, we showed in Section~\ref{sec:arepoct} that the presently implemented vector field constrained transport implementation {\ArepoCT} \citep{Mocz2016} is numerically unstable at high resolution for this problem and discussed the underlying reasons for it. This version should therefore not be used in production calculations. 

We note that, at comparable resolution, the finite volume fixed grid codes ({\AthenaCT} and {\RamsesCT}) are more than an order of magnitude faster when running this problem than the moving mesh code {\Arepo} and our DG code, independently of the choice of magnetic field implementation in the latter two. We caution that this is a qualitative comparison only, and that comparing the performance of schemes will always only be valid for specific problem and the properties of the specific machine used, and likely also depend on detailed choices of, for example, parallel decomposition.

Assuming that the amplification rate is a proxy for the numerical dissipation of the schemes, it is interesting to compare the runtime of codes at resolutions that achieve similar amplification rates. As an example, the amplification rate of {\AthenaCT} at $256^3$ ($0.94$) is slightly slower than the amplification rate of {\ArepoPowell} at $128^3$ ($1.18$), but {\ArepoPowell} at $128^3$ is $50\%$ more expensive to run to $t=100$ than {\AthenaCT} at $256^3$. The amplification rates of {\AthenaCT} and {\DGDedner} at $256^3$ resolution are similar ($0.94$ versus $0.90$), but at this resolution {\DGDedner} is almost $20\times$ more expensive to run to $t=100$ than {\AthenaCT}.

Our study focuses on a very specific setup of driven subsonic turbulence with purely numerical dissipation of the momentum and magnetic field. One obvious next step is to extend it to supersonic turbulence. Moreover, it would be interesting to implement explicit viscosity and resistivity in all codes and then compare them again in a similar setup of driven turbulence, but now with the same Reynolds and Prandtl numbers to confirm that they then obtain the same results. One could also use a set of high resolution simulations with known controlled viscosity and resistivity to compare the simulations with purely numerical viscosity and resistivity to better understand these effective numerical properties and how they differ between different codes. Finally, it would be interesting to extend the comparison to a wider range of numerical schemes, for example by including particle based schemes, or (numerically stable) vector field constrained transport schemes.

\section*{Acknowledgements}

The authors thank Jim Stone for important input and helpful discussions.
This work used the DiRAC Data Intensive service (CSD3) at the University of Cambridge, managed by the University of Cambridge University Information Services on behalf of the STFC DiRAC HPC Facility (\url{www.dirac.ac.uk}).
The DiRAC component of CSD3 at Cambridge was funded by BEIS, UKRI and STFC capital funding and STFC operations grants. DiRAC is part of the UKRI Digital Research Infrastructure. The authors would also like to acknowledge the use of the University of Exeter's Advanced Research Computing facilities (\texttt{isca}) in carrying out this work.
RB is supported by the SNSF through the Ambizione Grant PZ00P2\_223532.
CP acknowledges support from the European Research Council via the ERC Advanced Grant ``PICOGAL'' (project ID 101019746).
 
\section*{Data Availability}

The data underlying this article will be shared on reasonable request
to the corresponding author.

\bibliographystyle{mnras}

\begin{thebibliography}{}
\makeatletter
\relax
\def\mn@urlcharsother{\let\do\@makeother \do\$\do\&\do\#\do\^\do\_\do\%\do\~}
\def\mn@doi{\begingroup\mn@urlcharsother \@ifnextchar [ {\mn@doi@}
  {\mn@doi@[]}}
\def\mn@doi@[#1]#2{\def\@tempa{#1}\ifx\@tempa\@empty \href
  {http://dx.doi.org/#2} {doi:#2}\else \href {http://dx.doi.org/#2} {#1}\fi
  \endgroup}
\def\mn@eprint#1#2{\mn@eprint@#1:#2::\@nil}
\def\mn@eprint@arXiv#1{\href {http://arxiv.org/abs/#1} {{\tt arXiv:#1}}}
\def\mn@eprint@dblp#1{\href {http://dblp.uni-trier.de/rec/bibtex/#1.xml}
  {dblp:#1}}
\def\mn@eprint@#1:#2:#3:#4\@nil{\def\@tempa {#1}\def\@tempb {#2}\def\@tempc
  {#3}\ifx \@tempc \@empty \let \@tempc \@tempb \let \@tempb \@tempa \fi \ifx
  \@tempb \@empty \def\@tempb {arXiv}\fi \@ifundefined
  {mn@eprint@\@tempb}{\@tempb:\@tempc}{\expandafter \expandafter \csname
  mn@eprint@\@tempb\endcsname \expandafter{\@tempc}}}

\bibitem[\protect\citeauthoryear{{Balsara}}{{Balsara}}{2012}]{balsara_2012}
{Balsara} D.~S.,  2012, \mn@doi [Journal of Computational Physics]
  {10.1016/j.jcp.2011.12.025}, \href
  {https://ui.adsabs.harvard.edu/abs/2012JCoPh.231.7476B} {231, 7476}

\bibitem[\protect\citeauthoryear{Bauer \& Springel}{Bauer \&
  Springel}{2012}]{bauer_subsonic_2012}
Bauer A.,  Springel V.,  2012, \mn@doi [Monthly Notices of the Royal
  Astronomical Society] {10.1111/j.1365-2966.2012.21058.x}, 423, 2558

\bibitem[\protect\citeauthoryear{{Beattie}, {Federrath}, {Kriel}, {Mocz}  \&
  {Seta}}{{Beattie} et~al.}{2023}]{Beattie2023}
{Beattie} J.~R.,  {Federrath} C.,  {Kriel} N.,  {Mocz} P.,   {Seta} A.,  2023,
  \mn@doi [\mnras] {10.1093/mnras/stad1863}, \href
  {https://ui.adsabs.harvard.edu/abs/2023MNRAS.524.3201B} {524, 3201}

\bibitem[\protect\citeauthoryear{Brandenburg}{Brandenburg}{2011}]{Brandenburg2011}
Brandenburg A.,  2011, \mn@doi [The Astrophysical Journal]
  {10.1088/0004-637X/741/2/92}, 741, 92

\bibitem[\protect\citeauthoryear{{Brandenburg} \& {Subramanian}}{{Brandenburg}
  \& {Subramanian}}{2005}]{Brandenburg2005}
{Brandenburg} A.,  {Subramanian} K.,  2005, \mn@doi [\physrep]
  {10.1016/j.physrep.2005.06.005}, \href
  {https://ui.adsabs.harvard.edu/abs/2005PhR...417....1B} {417, 1}

\bibitem[\protect\citeauthoryear{{Brandenburg}, {Sokoloff}  \&
  {Subramanian}}{{Brandenburg} et~al.}{2012}]{Brandenburg2012}
{Brandenburg} A.,  {Sokoloff} D.,   {Subramanian} K.,  2012, \mn@doi [\ssr]
  {10.1007/s11214-012-9909-x}, \href
  {https://ui.adsabs.harvard.edu/abs/2012SSRv..169..123B} {169, 123}

\bibitem[\protect\citeauthoryear{{Cernetic}, {Springel}, {Guillet}  \&
  {Pakmor}}{{Cernetic} et~al.}{2023}]{cernetic2023}
{Cernetic} M.,  {Springel} V.,  {Guillet} T.,   {Pakmor} R.,  2023, \mn@doi
  [\mnras] {10.1093/mnras/stad1043}, \href
  {https://ui.adsabs.harvard.edu/abs/2023MNRAS.522..982C} {522, 982}

\bibitem[\protect\citeauthoryear{{Cernetic}, {Springel}, {Guillet}  \&
  {Pakmor}}{{Cernetic} et~al.}{2024}]{cernetic2024}
{Cernetic} M.,  {Springel} V.,  {Guillet} T.,   {Pakmor} R.,  2024, \mn@doi
  [\mnras] {10.1093/mnras/stae2192}, \href
  {https://ui.adsabs.harvard.edu/abs/2024MNRAS.534.1963C} {534, 1963}

\bibitem[\protect\citeauthoryear{{Colella} \& {Woodward}}{{Colella} \&
  {Woodward}}{1984}]{Colella1984}
{Colella} P.,  {Woodward} P.~R.,  1984, \mn@doi [Journal of Computational
  Physics] {10.1016/0021-9991(84)90143-8}, \href
  {https://ui.adsabs.harvard.edu/abs/1984JCoPh..54..174C} {54, 174}

\bibitem[\protect\citeauthoryear{{Crain} \& {van de Voort}}{{Crain} \& {van de
  Voort}}{2023}]{Crain2023}
{Crain} R.~A.,  {van de Voort} F.,  2023, \mn@doi [\araa]
  {10.1146/annurev-astro-041923-043618}, \href
  {https://ui.adsabs.harvard.edu/abs/2023ARA&A..61..473C} {61, 473}

\bibitem[\protect\citeauthoryear{{Dedner}, {Kemm}, {Kr{\"o}ner}, {Munz},
  {Schnitzer}  \& {Wesenberg}}{{Dedner} et~al.}{2002}]{dedner2002}
{Dedner} A.,  {Kemm} F.,  {Kr{\"o}ner} D.,  {Munz} C.-D.,  {Schnitzer} T.,
  {Wesenberg} M.,  2002, \mn@doi [Journal of Computational Physics]
  {10.1006/jcph.2001.6961}, \href
  {https://ui.adsabs.harvard.edu/abs/2002JCoPh.175..645D} {175, 645}

\bibitem[\protect\citeauthoryear{{Elmegreen} \& {Scalo}}{{Elmegreen} \&
  {Scalo}}{2004}]{Elmegreen2004}
{Elmegreen} B.~G.,  {Scalo} J.,  2004, \mn@doi [\araa]
  {10.1146/annurev.astro.41.011802.094859}, \href
  {https://ui.adsabs.harvard.edu/abs/2004ARA&A..42..211E} {42, 211}

\bibitem[\protect\citeauthoryear{{Evans} \& {Hawley}}{{Evans} \&
  {Hawley}}{1988}]{Evans1988}
{Evans} C.~R.,  {Hawley} J.~F.,  1988, \mn@doi [\apj] {10.1086/166684}, \href
  {https://ui.adsabs.harvard.edu/abs/1988ApJ...332..659E} {332, 659}

\bibitem[\protect\citeauthoryear{Federrath, {Roman-Duval}, Klessen, Schmidt  \&
  Mac~Low}{Federrath
  et~al.}{2010}]{federrath_comparingstatisticsinterstellar_2010}
Federrath C.,  {Roman-Duval} J.,  Klessen R.~S.,  Schmidt W.,   Mac~Low M.-M.,
  2010, \mn@doi [Astronomy and Astrophysics] {10.1051/0004-6361/200912437},
  512, A81

\bibitem[\protect\citeauthoryear{{Feldmann} \& {Bieri}}{{Feldmann} \&
  {Bieri}}{2026}]{Feldmann2026}
{Feldmann} R.,  {Bieri} R.,  2026, in Encyclopedia of Astrophysics, Volume 4.
  pp 576--599 (\mn@eprint {arXiv} {2507.08925}),
  \mn@doi{10.1016/B978-0-443-21439-4.00111-5}

\bibitem[\protect\citeauthoryear{Gottlieb}{Gottlieb}{2005}]{gottlieb2005}
Gottlieb S.,  2005, \mn@doi [Journal of Scientific Computing]
  {10.1007/s10915-004-4635-5}, 25, 105

\bibitem[\protect\citeauthoryear{{Grete}, {O'Shea}  \& {Beckwith}}{{Grete}
  et~al.}{2023}]{grete_2023}
{Grete} P.,  {O'Shea} B.~W.,   {Beckwith} K.,  2023, \mn@doi [\apjl]
  {10.3847/2041-8213/acaea7}, \href
  {https://ui.adsabs.harvard.edu/abs/2023ApJ...942L..34G} {942, L34}

\bibitem[\protect\citeauthoryear{{Guillet}, {Pakmor}, {Springel},
  {Chandrashekar}  \& {Klingenberg}}{{Guillet} et~al.}{2019}]{guillet2019}
{Guillet} T.,  {Pakmor} R.,  {Springel} V.,  {Chandrashekar} P.,
  {Klingenberg} C.,  2019, \mn@doi [\mnras] {10.1093/mnras/stz314}, \href
  {https://ui.adsabs.harvard.edu/abs/2019MNRAS.485.4209G} {485, 4209}

\bibitem[\protect\citeauthoryear{{Kazantsev}, {Ruzmaikin}  \&
  {Sokolov}}{{Kazantsev} et~al.}{1985}]{kazantsev1985}
{Kazantsev} A.~P.,  {Ruzmaikin} A.~A.,   {Sokolov} D.~D.,  1985, Zhurnal
  Eksperimentalnoi i Teoreticheskoi Fiziki, \href
  {https://ui.adsabs.harvard.edu/abs/1985ZhETF..88..487K} {88, 487}

\bibitem[\protect\citeauthoryear{{Kriel}, {Beattie}, {Seta}  \&
  {Federrath}}{{Kriel} et~al.}{2022}]{Kriel2022}
{Kriel} N.,  {Beattie} J.~R.,  {Seta} A.,   {Federrath} C.,  2022, \mn@doi
  [\mnras] {10.1093/mnras/stac969}, \href
  {https://ui.adsabs.harvard.edu/abs/2022MNRAS.513.2457K} {513, 2457}

\bibitem[\protect\citeauthoryear{Kritsuk et~al.,}{Kritsuk
  et~al.}{2011}]{kritsuk_comparing_2011}
Kritsuk A.~G.,  et~al., 2011, \mn@doi [The Astrophysical Journal]
  {10.1088/0004-637X/737/1/13}, 737, 13

\bibitem[\protect\citeauthoryear{{Martin-Alvarez}, {Devriendt}, {Slyz}  \&
  {Teyssier}}{{Martin-Alvarez} et~al.}{2018}]{MartinAlvarez2018}
{Martin-Alvarez} S.,  {Devriendt} J.,  {Slyz} A.,   {Teyssier} R.,  2018,
  \mn@doi [\mnras] {10.1093/mnras/sty1623}, \href
  {https://ui.adsabs.harvard.edu/abs/2018MNRAS.479.3343M} {479, 3343}

\bibitem[\protect\citeauthoryear{{Miyoshi} \& {Kusano}}{{Miyoshi} \&
  {Kusano}}{2005}]{miyoshi2005}
{Miyoshi} T.,  {Kusano} K.,  2005, \mn@doi [Journal of Computational Physics]
  {10.1016/j.jcp.2005.02.017}, \href
  {https://ui.adsabs.harvard.edu/abs/2005JCoPh.208..315M} {208, 315}

\bibitem[\protect\citeauthoryear{{Mocz}, {Pakmor}, {Springel}, {Vogelsberger},
  {Marinacci}  \& {Hernquist}}{{Mocz} et~al.}{2016}]{Mocz2016}
{Mocz} P.,  {Pakmor} R.,  {Springel} V.,  {Vogelsberger} M.,  {Marinacci} F.,
  {Hernquist} L.,  2016, \mn@doi [\mnras] {10.1093/mnras/stw2004}, \href
  {https://ui.adsabs.harvard.edu/abs/2016MNRAS.463..477M} {463, 477}

\bibitem[\protect\citeauthoryear{{Pakmor} \& {Springel}}{{Pakmor} \&
  {Springel}}{2013}]{pakmor2013}
{Pakmor} R.,  {Springel} V.,  2013, \mn@doi [\mnras] {10.1093/mnras/stt428},
  \href {https://ui.adsabs.harvard.edu/abs/2013MNRAS.432..176P} {432, 176}

\bibitem[\protect\citeauthoryear{{Pakmor}, {Bauer}  \& {Springel}}{{Pakmor}
  et~al.}{2011}]{pakmor2011}
{Pakmor} R.,  {Bauer} A.,   {Springel} V.,  2011, \mn@doi [\mnras]
  {10.1111/j.1365-2966.2011.19591.x}, \href
  {https://ui.adsabs.harvard.edu/abs/2011MNRAS.418.1392P} {418, 1392}

\bibitem[\protect\citeauthoryear{{Pakmor}, {Springel}, {Bauer}, {Mocz},
  {Munoz}, {Ohlmann}, {Schaal}  \& {Zhu}}{{Pakmor} et~al.}{2016}]{pakmor2016}
{Pakmor} R.,  {Springel} V.,  {Bauer} A.,  {Mocz} P.,  {Munoz} D.~J.,
  {Ohlmann} S.~T.,  {Schaal} K.,   {Zhu} C.,  2016, \mn@doi [\mnras]
  {10.1093/mnras/stv2380}, \href
  {https://ui.adsabs.harvard.edu/abs/2016MNRAS.455.1134P} {455, 1134}

\bibitem[\protect\citeauthoryear{{Pakmor} et~al.,}{{Pakmor}
  et~al.}{2017}]{Pakmor2017}
{Pakmor} R.,  et~al., 2017, \mn@doi [\mnras] {10.1093/mnras/stx1074}, \href
  {https://ui.adsabs.harvard.edu/abs/2017MNRAS.469.3185P} {469, 3185}

\bibitem[\protect\citeauthoryear{{Pakmor} et~al.,}{{Pakmor}
  et~al.}{2024}]{Pakmor2024}
{Pakmor} R.,  et~al., 2024, \mn@doi [\mnras] {10.1093/mnras/stae112}, \href
  {https://ui.adsabs.harvard.edu/abs/2024MNRAS.528.2308P} {528, 2308}

\bibitem[\protect\citeauthoryear{{Pfrommer}, {Werhahn}, {Pakmor}, {Girichidis}
  \& {Simpson}}{{Pfrommer} et~al.}{2022}]{Pfrommer2022}
{Pfrommer} C.,  {Werhahn} M.,  {Pakmor} R.,  {Girichidis} P.,   {Simpson}
  C.~M.,  2022, \mn@doi [\mnras] {10.1093/mnras/stac1808}, \href
  {https://ui.adsabs.harvard.edu/abs/2022MNRAS.515.4229P} {515, 4229}

\bibitem[\protect\citeauthoryear{{Powell}, {Roe}, {Linde}, {Gombosi}  \& {De
  Zeeuw}}{{Powell} et~al.}{1999}]{powell1999}
{Powell} K.~G.,  {Roe} P.~L.,  {Linde} T.~J.,  {Gombosi} T.~I.,   {De Zeeuw}
  D.~L.,  1999, \mn@doi [Journal of Computational Physics]
  {10.1006/jcph.1999.6299}, \href
  {https://ui.adsabs.harvard.edu/abs/1999JCoPh.154..284P} {154, 284}

\bibitem[\protect\citeauthoryear{{Rieder} \& {Teyssier}}{{Rieder} \&
  {Teyssier}}{2016}]{Rieder2016}
{Rieder} M.,  {Teyssier} R.,  2016, \mn@doi [\mnras] {10.1093/mnras/stv2985},
  \href {https://ui.adsabs.harvard.edu/abs/2016MNRAS.457.1722R} {457, 1722}

\bibitem[\protect\citeauthoryear{{Rieder} \& {Teyssier}}{{Rieder} \&
  {Teyssier}}{2017a}]{Rieder2017}
{Rieder} M.,  {Teyssier} R.,  2017a, \mn@doi [\mnras] {10.1093/mnras/stx1670},
  \href {https://ui.adsabs.harvard.edu/abs/2017MNRAS.471.2674R} {471, 2674}

\bibitem[\protect\citeauthoryear{{Rieder} \& {Teyssier}}{{Rieder} \&
  {Teyssier}}{2017b}]{Rieder2017b}
{Rieder} M.,  {Teyssier} R.,  2017b, \mn@doi [\mnras] {10.1093/mnras/stx2276},
  \href {https://ui.adsabs.harvard.edu/abs/2017MNRAS.472.4368R} {472, 4368}

\bibitem[\protect\citeauthoryear{{Schaal}, {Bauer}, {Chandrashekar}, {Pakmor},
  {Klingenberg}  \& {Springel}}{{Schaal} et~al.}{2015}]{schaal2015}
{Schaal} K.,  {Bauer} A.,  {Chandrashekar} P.,  {Pakmor} R.,  {Klingenberg} C.,
    {Springel} V.,  2015, \mn@doi [\mnras] {10.1093/mnras/stv1859}, \href
  {https://ui.adsabs.harvard.edu/abs/2015MNRAS.453.4278S} {453, 4278}

\bibitem[\protect\citeauthoryear{{Schekochihin}, {Cowley}, {Taylor}, {Maron}
  \& {McWilliams}}{{Schekochihin} et~al.}{2004}]{Schekochihin2004}
{Schekochihin} A.~A.,  {Cowley} S.~C.,  {Taylor} S.~F.,  {Maron} J.~L.,
  {McWilliams} J.~C.,  2004, \mn@doi [\apj] {10.1086/422547}, \href
  {https://ui.adsabs.harvard.edu/abs/2004ApJ...612..276S} {612, 276}

\bibitem[\protect\citeauthoryear{Schmidt, Hillebrandt  \& Niemeyer}{Schmidt
  et~al.}{2006}]{schmidtNumericalDissipationBottleneck2006}
Schmidt W.,  Hillebrandt W.,   Niemeyer J.~C.,  2006, \mn@doi [Computers \&
  Fluids] {10.1016/j.compfluid.2005.03.002}, 35, 353

\bibitem[\protect\citeauthoryear{{Springel}}{{Springel}}{2010}]{springel2010}
{Springel} V.,  2010, \mn@doi [\mnras] {10.1111/j.1365-2966.2009.15715.x},
  \href {https://ui.adsabs.harvard.edu/abs/2010MNRAS.401..791S} {401, 791}

\bibitem[\protect\citeauthoryear{{Stone}, {Gardiner}, {Teuben}, {Hawley}  \&
  {Simon}}{{Stone} et~al.}{2008}]{stone2008}
{Stone} J.~M.,  {Gardiner} T.~A.,  {Teuben} P.,  {Hawley} J.~F.,   {Simon}
  J.~B.,  2008, \mn@doi [\apjs] {10.1086/588755}, \href
  {https://ui.adsabs.harvard.edu/abs/2008ApJS..178..137S} {178, 137}

\bibitem[\protect\citeauthoryear{{Stone}, {Tomida}, {White}  \&
  {Felker}}{{Stone} et~al.}{2020}]{stone2020}
{Stone} J.~M.,  {Tomida} K.,  {White} C.~J.,   {Felker} K.~G.,  2020, \mn@doi
  [\apjs] {10.3847/1538-4365/ab929b}, \href
  {https://ui.adsabs.harvard.edu/abs/2020ApJS..249....4S} {249, 4}

\bibitem[\protect\citeauthoryear{{Teyssier}}{{Teyssier}}{2002}]{teyssier2002}
{Teyssier} R.,  2002, \mn@doi [\aap] {10.1051/0004-6361:20011817}, \href
  {https://ui.adsabs.harvard.edu/abs/2002A&A...385..337T} {385, 337}

\bibitem[\protect\citeauthoryear{{Teyssier}, {Fromang}  \& {Dormy}}{{Teyssier}
  et~al.}{2006}]{teyssier2006}
{Teyssier} R.,  {Fromang} S.,   {Dormy} E.,  2006, \mn@doi [Journal of
  Computational Physics] {10.1016/j.jcp.2006.01.042}, \href
  {https://ui.adsabs.harvard.edu/abs/2006JCoPh.218...44T} {218, 44}

\bibitem[\protect\citeauthoryear{{Tomida}, {Sadanari}, {Takasao}  \&
  {Iwasaki}}{{Tomida} et~al.}{2026}]{Tomida2026}
{Tomida} K.,  {Sadanari} K.~E.,  {Takasao} S.,   {Iwasaki} K.,  2026, \mn@doi
  [arXiv e-prints] {10.48550/arXiv.2605.07928}, \href
  {https://ui.adsabs.harvard.edu/abs/2026arXiv260507928T} {p. arXiv:2605.07928}

\bibitem[\protect\citeauthoryear{{Tumlinson}, {Peeples}  \& {Werk}}{{Tumlinson}
  et~al.}{2017}]{Tumlinson2017}
{Tumlinson} J.,  {Peeples} M.~S.,   {Werk} J.~K.,  2017, \mn@doi [\araa]
  {10.1146/annurev-astro-091916-055240}, \href
  {https://ui.adsabs.harvard.edu/abs/2017ARA&A..55..389T} {55, 389}

\bibitem[\protect\citeauthoryear{{Vogelsberger}, {Sijacki}, {Kere{\v{s}}},
  {Springel}  \& {Hernquist}}{{Vogelsberger} et~al.}{2012}]{Vogelsberger2012}
{Vogelsberger} M.,  {Sijacki} D.,  {Kere{\v{s}}} D.,  {Springel} V.,
  {Hernquist} L.,  2012, \mn@doi [\mnras] {10.1111/j.1365-2966.2012.21590.x},
  \href {https://ui.adsabs.harvard.edu/abs/2012MNRAS.425.3024V} {425, 3024}

\bibitem[\protect\citeauthoryear{{Vogelsberger}, {Marinacci}, {Torrey}  \&
  {Puchwein}}{{Vogelsberger} et~al.}{2020}]{Vogelsberger2020}
{Vogelsberger} M.,  {Marinacci} F.,  {Torrey} P.,   {Puchwein} E.,  2020,
  \mn@doi [Nature Reviews Physics] {10.1038/s42254-019-0127-2}, \href
  {https://ui.adsabs.harvard.edu/abs/2020NatRP...2...42V} {2, 42}

\bibitem[\protect\citeauthoryear{{Weinberger}, {Springel}  \&
  {Pakmor}}{{Weinberger} et~al.}{2020}]{weinberger2020}
{Weinberger} R.,  {Springel} V.,   {Pakmor} R.,  2020, \mn@doi [\apjs]
  {10.3847/1538-4365/ab908c}, \href
  {https://ui.adsabs.harvard.edu/abs/2020ApJS..248...32W} {248, 32}

\bibitem[\protect\citeauthoryear{{Wissing} \& {Shen}}{{Wissing} \&
  {Shen}}{2023}]{Wissing2023}
{Wissing} R.,  {Shen} S.,  2023, \mn@doi [\aap] {10.1051/0004-6361/202244753},
  \href {https://ui.adsabs.harvard.edu/abs/2023A&A...673A..47W} {673, A47}

\makeatother
\end{thebibliography}


\bsp	
\label{lastpage}
\end{document}